\documentclass[11pt]{article}

\usepackage[linesnumbered,lined,boxed,commentsnumbered]{algorithm2e}

\usepackage{arxiv}

\usepackage{epsfig}
\usepackage{ulem}
\usepackage[letterpaper]{geometry}
\usepackage{multirow}
\usepackage{graphicx}
\usepackage[usenames,dvipsnames]{color}
\usepackage{bbm}

\usepackage{arxiv}

\usepackage{amsthm}
\usepackage{amssymb}
\usepackage[utf8]{inputenc}
\usepackage{amsmath}
\usepackage{amsfonts}
\usepackage{graphicx}
\usepackage[colorinlistoftodos]{todonotes}
\usepackage{thmtools}
\usepackage{fullpage}
\usepackage{booktabs}
\usepackage{bbm}
\usepackage{hyperref}

\newcommand{\Pc}{\mathcal{P}}

\newcommand{\A}{\mathbf{A}}

\newcommand{\vol}{\textrm{vol}}
\newcommand{\ABCD}{\textbf{ABCD}}

\newcommand{\N}{{\mathbb N}}
\newcommand{\R}{{\mathbb R}}
\newcommand{\E}{\mathbb E}

\theoremstyle{plain}

\title{Artificial Benchmark for Community Detection with Outliers (ABCD+o)}

\author{
Bogumi\l{} Kami\'nski\thanks{Decision Analysis and Support Unit, SGH Warsaw School of Economics, Warsaw, Poland; e-mail: \texttt{bogumil.kaminski@sgh.waw.pl}}
\And
Pawe\l{}~Pra\l{}at\thanks{Department of Mathematics, Toronto Metropolitan University, Toronto, ON, Canada; e-mail: \texttt{pralat@ryerson.ca}. Part of this work was done while the author was visiting the Simons Institute for the Theory of Computing.}
\And
Fran\c{c}ois Th\'eberge\thanks{Tutte Institute for Mathematics and Computing, Ottawa, ON, Canada; email: \texttt{theberge@ieee.org}}
}

\begin{document}
\maketitle              

\begin{abstract}
The \textbf{A}rtificial \textbf{B}enchmark for \textbf{C}ommunity \textbf{D}etection graph (\ABCD) is a random graph model with community structure and power-law distribution for both degrees and community sizes. The model generates graphs with similar properties as the well-known \textbf{LFR} one, and its main parameter $\xi$ can be tuned to mimic its counterpart in the \textbf{LFR} model, the mixing parameter $\mu$. 

In this paper, we extend the \ABCD\ model to include potential outliers. We perform some exploratory experiments on both the new \textbf{ABCD+o} model as well as a real-world network to show that outliers pose some distinguishable properties. This ensures that our new model may serve as a benchmark of outlier detection algorithms.
\end{abstract}

\section{Introduction}

One of the most important features of real-world networks is their community structure, as it reveals the internal organization of nodes~\cite{fortunato2010community}. In social networks, communities may represent groups by interest; in citation networks, they correspond to related papers; on the Web, communities are formed by pages on related topics, etc. Being able to identify communities in a network could help us to exploit this network more effectively. Grouping like-minded users or similar-looking items together is important for a wide range of applications, including controlling epidemics~\cite{ghalmane2019immunization}, recommendation systems, anomaly or outlier detection, fraud detection, rumor or fake news detection, etc.~\cite{javed2018community}. There is also growing literature introducing community-aware centrality measures that exploit both local and global properties of networks~\cite{curado2023novel,rajeh2023comparative}. For more discussion around various aspects of mining complex networks, see for example,~\cite{Newman2018,kaminski2021mining}.

One of the major current challenges in detecting communities is that most of the existing algorithms treat all nodes the same way. That is, they try to assign them to precisely one community. On the other hand, many complex networks (regardless of whether their nodes correspond to, say, users of some social media or movies on Netflix) consist of nodes that are active participants of their own communities while others are not~\cite{liu2020gmm}. As a result, there is a need to detect nodes that are not a strong part of any of the communities. In this paper we refer to such nodes as \emph{outliers}.

Another feature of empirical graphs is that some communities might be overlapping, which is reflected by some of the nodes belonging to a few communities via fuzzy membership. For example, a label propagation method~\cite{gregory2010finding} and a non-negative matrix factorization approach~\cite{yang2013overlapping} are introduced to find overlapping communities. Independently, a \textbf{Louvain}-based algorithm is proposed in~\cite{singh2021ni} to detect overlapping communities with influence analysis.

The survey~\cite{akoglu2015graph} reviews various methods and approaches to graph anomaly detection. In particular, Section~2.1.2 in this survey contains a review of various methods that are used for identifying community-based outliers. This type of outlier is a subject of interest in this paper.

Selected relevant approaches related to community-based outliers detection can be found in~\cite{bandyopadhyay2020integrating,Chakrabarti2004,gaucher2021outlier,Liu2015,sun2005,Viswanath2010}, but more research is expected to be pursued in the near future. There are two reasons supporting such observation. Firstly, in many applications, outliers themselves are objects of interest. Secondly, proper handling of graph outliers when mining complex networks is, in our opinion, more important and more sophisticated than in standard machine learning when working with tabular data. Indeed, many procedures used in mining complex networks (e.g.,\ graph embeddings) are affected by the presence of outliers. However, one cannot simply remove them from the data, as opposed to standard machine learning, where such a procedure is sometimes applied. The issue is that removing nodes in networks affects the properties of other nodes and changes the underlying graph's structure and topology.

\medskip

Another well-known challenge recognized by researchers is that there is a limited number of datasets with ground truth identified and labeled. As a result, there is a need for synthetic random graph models with community structure that resemble real-world networks in order to benchmark algorithms that aim to analyze graph community structure. The \textbf{LFR} (Lancichinetti, Fortunato, Radicchi) model~\cite{lancichinetti2009benchmarks,lancichinetti2008benchmark} generates networks with communities, and at the same time, it allows for the heterogeneity in the distributions of both node degrees and of community sizes. It became a standard and extensively used method for generating artificial networks with (non-overlapping) community structure. 

Unfortunately, the situation is much more challenging if one needs a synthetic model with outliers. There seems to be no standard model that one may use. For example, in~\cite{gaucher2021outlier} the authors adjust the classical Stochastic Block Model to simultaneously take into account the community structure and outliers by introducing different probabilities of connection between inliers and pairs involving outliers. To validate algorithms tested in~\cite{bandyopadhyay2020integrating}, the authors start with a synthetic \textbf{LFR} network or a real-world one and then randomly perturb edges around some randomly selected nodes in order to create artificial outliers. \textbf{LFR} itself~\cite{lancichinetti2009benchmarks} has some basic functionality to create overlapping clusters but not outliers. 

\medskip

This paper is an extended version of the proceeding paper~\cite{kaminski2022outliers} in which we revisit the Artificial Benchmark for Community Detection (\textbf{ABCD}) graph~\cite{kaminski2021artificial}. This model was recently introduced and implemented\footnote{\url{https://github.com/bkamins/ABCDGraphGenerator.jl/}}, including a fast implementation that uses multiple threads (\textbf{ABCDe})~\cite{kaminski2022fast}\footnote{\url{https://github.com/tolcz/ABCDeGraphGenerator.jl/}}. Undirected variant of \textbf{LFR} and \textbf{ABCD} produce synthetic networks with comparable properties but \textbf{ABCD}/\textbf{ABCDe} is significantly faster than \textbf{LFR} and can be easily tuned to allow the user to make a smooth transition between the two extremes: pure (disjoint) communities and random graph with no community structure. Moreover, it is easier to analyze theoretically. For example, various theoretical asymptotic properties of the \ABCD\ model are analyzed in~\cite{kaminski2022modularity}, including the modularity function that is, arguably, the most important graph property of networks in the context of community detection.

In this paper, we extend the original \textbf{ABCD} model to include potential outliers, \textbf{ABCD+o} model (see Section~\ref{sec:model}). We examine one of the few real-world networks with identified outliers, the College Football Graph (see Subsection~\ref{sec:football}), and identify distinctive properties of outliers that are present in this network. We then perform simulations with our new \textbf{ABCD+o} model to show that its outliers possess similar properties (see Subsections~\ref{sec:participation},~\ref{sec:ECG},~\ref{sec:edges}, and~\ref{sec:entropy}). This validates that our model mimics real-world networks with the presence of outliers and so may serve as a benchmark of outlier detection algorithms. Additionally, to illustrate the need for proper outlier-detection algorithms, we show in Subsection~\ref{sec:properties} that some classical and widely used centrality measures fail to distinguish outliers from regular nodes both for College Football Graph and \textbf{ABCD+o} model.

The applications presented in Section~\ref{sec:experiments} show usefulness of the proposed \textbf{ABCD+o} model as a benchmark for such tests. In particular, we show that, as opposed to real-world graphs (which have a fixed structure), we can analyze the impact of varying graph parameters, such as the average degree or strength of communities, on the process of outlier detection.

Future directions are briefly mentioned in Section~\ref{sec:future}. 
The associated Jupyter notebook can be found on GitHub repository\footnote{\url{https://github.com/ftheberge/ABCDoExperiments}}.

\section{Adjusting the \ABCD\ Model to Include Outliers}\label{sec:model}

We start this section with a brief description of the \ABCD\ model taken from~\cite{kaminski2022fast}; details can be found in~\cite{kaminski2021artificial} or in~\cite{kaminski2022modularity}. We then carefully explain the adjustments needed to incorporate the existence of outliers.

\subsection{The Original Model}

As in the {\bf LFR} model~\cite{lancichinetti2008benchmark,lancichinetti2009benchmarks}, for a given number of nodes $n$, we start by generating a power law distribution both for the degrees and community sizes. Those are governed by the power law exponent parameters $(\gamma,\beta)$. We also provide additional information to the model, again as it is done in {\bf LFR}, namely, the average and the maximum degree, and the range for the community sizes. The user may alternatively provide a specific degree distribution and/or community sizes.

For each community, we generate a random {\it community} subgraph on the nodes from a given community using either the \textbf{configuration model}~\cite{bollobas1980probabilistic} (see~\cite{bender1978asymptotic,wormald1984generating,wormald1999models} for related models and results) which preserves the exact degree distribution, or the \textbf{Chung-Lu model}~\cite{chung2006complex} which preserves the expected degree distribution. On top of it, we independently generate a {\it background} random graph on all the nodes. Everything is tuned so that the degree distribution of the union of all graphs follows the desired degree distribution (only in expectation in the case of the Chung-Lu variant). The mixing parameter $\xi$ guides the proportion of edges that are generated via the background graph. In particular, in the two extreme cases, when $\xi=1$ the graph has no community structure while if $\xi=0$, then we get disjoint communities. In order to generate simple graphs, we may have to do some re-sampling or edge re-wiring, which are described in~\cite{kaminski2021artificial}.

During this process, larger communities will additionally get some more internal edges due to the background graph. As argued in~\cite{kaminski2021artificial}, this ``global'' variant of the model is more natural and so we recommend it. However, in order to provide a variant where the expected proportion of internal edges is exactly the same for every community (as it is done in {\bf LFR}), we also provide a ``local'' variant of {\bf ABCD} in which the mixing parameter $\xi$ is automatically adjusted for every community.

\begin{figure}
\centering
\includegraphics[scale=0.32]{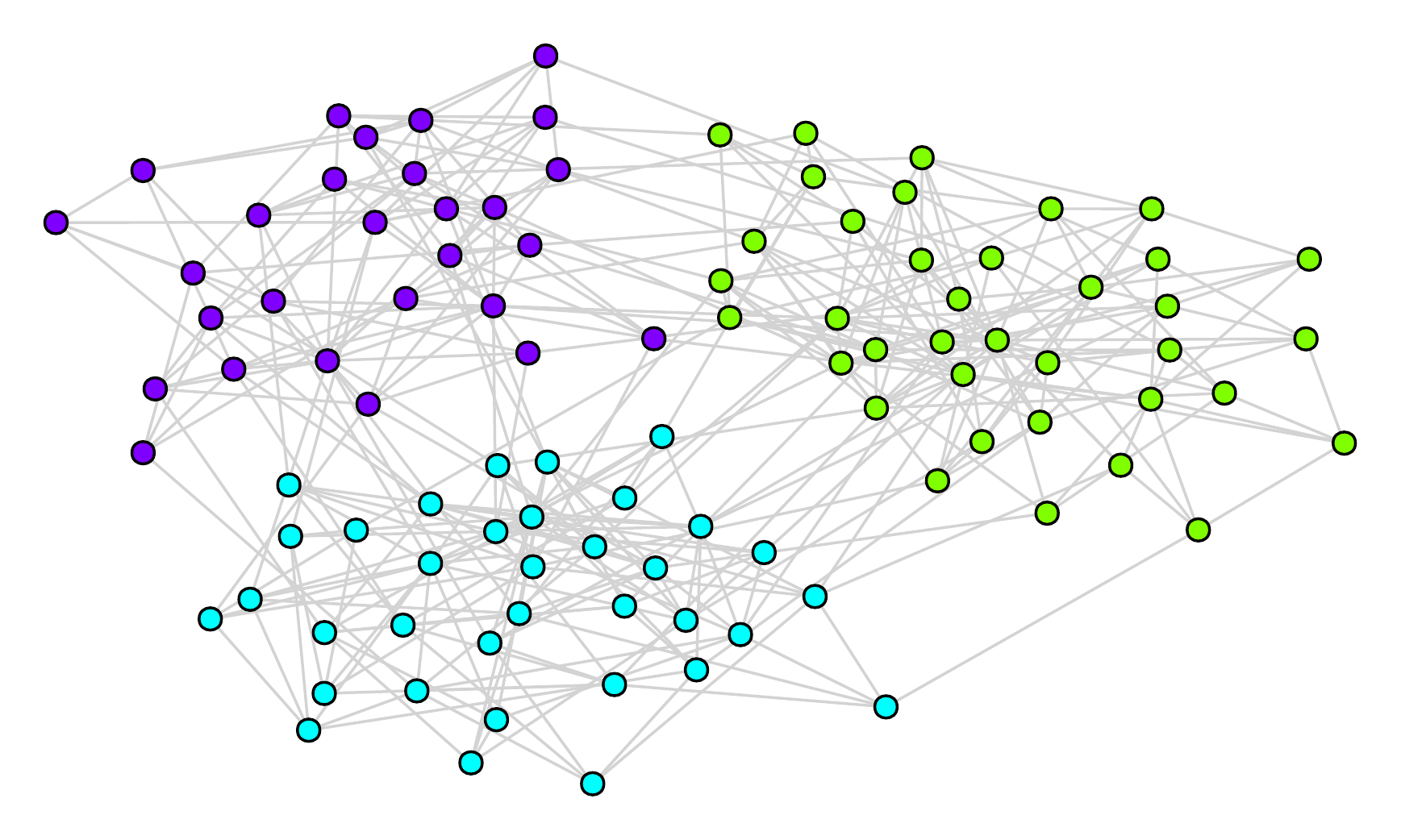} 
\hspace{.5cm}
\includegraphics[scale=0.32]{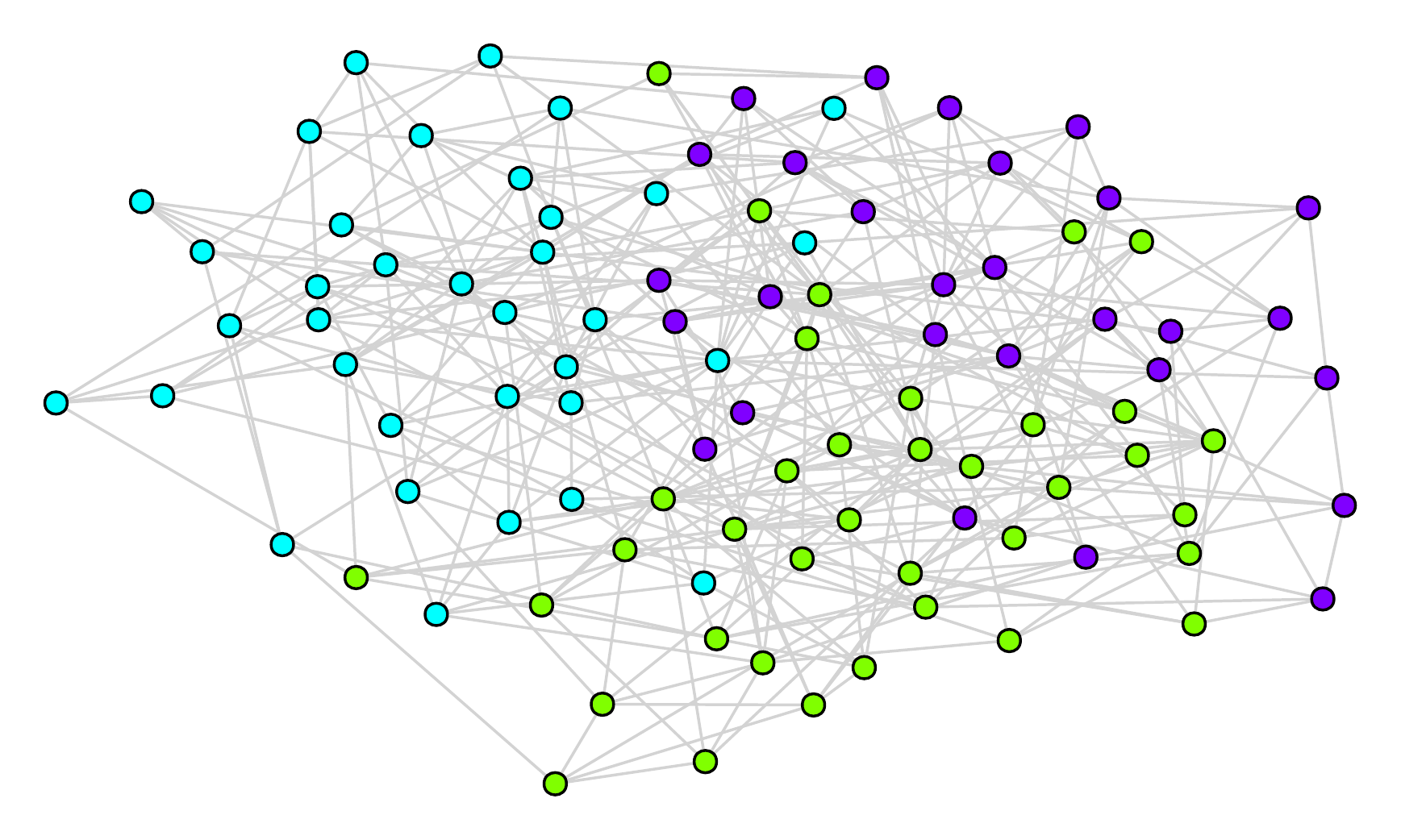} 
\caption{Two examples of \textbf{ABCD} graphs with low level of noise ($\xi=0.2$, left) and high level of noise ($\xi=0.4$, right).}
\label{fig:examples}
\end{figure}

Two examples of \textbf{ABCD} graphs on $n=100$ nodes are presented in Figure~\ref{fig:examples} (a standard Fruchterman-Reingold layout was used to plot them). Degree distribution was generated with power law exponent $\gamma=2.5$ with minimum and maximum values of 5 and 15, respectively. Community sizes were generated with power law exponent $\beta = 1.5$ with minimum and maximum values 20 and 40, respectively; communities are shown in different colors. The global variant and the configuration model were used to generate the graphs. The left plot has the mixing parameter set $\xi=0.2$ while the ``noisier'' graph on the right plot has the parameter fixed to $\xi=0.4$.

\subsection{Adjusting the Model to Include Outliers}

The adjusted model, \textbf{ABCD+o} (\ABCD\ with \textbf{outliers}), will have an additional parameter $s_0$ which is equal to the number of outliers. Because of a well-structured and flexible design of the original model, adjusting it to include outliers is simple. One trivial adjustment needed is the way the distribution of community sizes is generated. A slightly more delicate modification is needed in the process of assigning nodes to communities. However, before that, the algorithm needs to select suitable nodes for outliers. Below, we independently discuss these issues and explain how they are generalized.

The \textbf{ABCD+o} extension is defined only for the default settings of the original \textbf{ABCD} algorithm, namely, for the global version of the algorithm, configuration model used to generate community and background graphs, and accepts only parameter $\xi$ as the level of noise.
As in the original \ABCD\ model, the degree distribution is generated randomly following the (truncated) \emph{power-law distribution} $\Pc(\gamma, \delta, \Delta)$ with exponent $\gamma \in \R_+$, minimum value $\delta$, and maximum value $\Delta \ge \delta$. No adjustment is needed.

\subsubsection*{Distribution of Community Sizes.}
Let $\beta \in \R_+$, $s, S \in \N$ such that $\delta < s \le S$. Community sizes in the original \ABCD\ model are generated randomly following the (truncated) \emph{power-law distribution} $\Pc(\beta, s, S)$ with exponent $\beta$, minimum value $s$, and maximum value $S$. It is recommended to use $\beta \in (1,2)$, some relatively small value of $s$ such as 100 or 500, and $S$ larger than $\Delta$. The condition for $S$ is needed to make sure large degree nodes have large enough communities to be assigned to. Similarly, the assumption that $s \ge \delta+1$ is required to guarantee that small communities are not too small and in consequence that they can accommodate small degree nodes. These conditions are needed to make sure that generating a simple graph with the desired properties is feasible.

Communities in the original model are generated with this distribution as long as the sum of their sizes is less than $n$, the desired number of nodes. After drawing a predetermined number of samples from this distribution, the algorithm is selecting one sequence with a sum as close to $n$ as possible and carefully adjusts it, if needed. 

Since there are $s_0$ outliers in the new model (\textbf{ABCD+o}), the community sizes ($s_i$, $i \in [\ell] := \{1, \ldots, \ell\}$) are generated as in the original model but this time with the condition that the sum of their sizes is equal to $n-s_0$ (instead of $n$). 

\subsubsection*{Assigning Nodes to Outliers.}

Parameter $\xi \in (0,1)$ reflects the amount of noise in the network. It controls the fraction of edges that are between communities. Indeed, in the original \ABCD\ model, asymptotically (but not exactly) $1-\xi$ fraction of edges end up within one of the communities. Each node in the original model has its degree $w_i$ split into two parts: \emph{community degree} $y_i$ and \emph{background degree} $z_i$ ($w_i=y_i+z_i$). The goal is to get $y_i \approx (1-\xi) w_i$ and $z_i \approx \xi w_i$. However, both $y_i$ and $z_i$ have to be non-negative integers, and for each community $C \subseteq V$, $\sum_{i \in C} y_i$ has to be even. Fortunately, this can be easily achieved by an appropriate random rounding of $(1-\xi) w_i$ to the nearest integers. 

In the generalized \textbf{ABCD+o} model, each non-outlier has its degree $w_i$ split into $y_i$ and $z_i$, as in the original model. These nodes will be assigned to one community. On the other hand, the outlier nodes will not get assigned to any community so all of the incident edges will be generated in the background graph, thus the corresponding neighbors will be sampled from the entire graph. As a result, their degrees will satisfy $w_i = z_i$. Note that the only potential problem with outliers that might occur is when $\xi$ is close to zero. In the extreme case when $\xi=0$, only outliers have a non-zero degree in the background graph. In order to make sure that there exists a simple graph that satisfies the required degree distribution, in such extreme situations all outliers must have degrees smaller than $s_0$. The model needs to be prepared for such potential problems but in practice (when the number of nodes $n$ is large, the number of outliers $s_0$ is relatively small, and the level of noise $\xi$ is not zero) there are plenty of nodes with a non-zero degree in the background graph and so there is no restriction for outliers.

To prepare for potential problems we do the following. Once the degree of each node $w_i$ is split into $y_i$ and $z_i$, we get a lower bound for the number of nodes that will have a non-zero degree in the background graph, namely, 
$$
L := |\{ v \in V : z_i \ge 1 \}|. 
$$
Note that $\bar{L} = \E[L] =\sum_{i\in V}\min(1, \xi w_i)$ since each node with $\xi w_i \ge 1$ satisfies $z_i \ge 1$ and each node with $\xi w_i < 1$ has $z_i = 1$ with probability $\xi w_i$ and $z_i=0$ otherwise. Moreover, since by default outliers have $z_i = w_i \ge 1$, there will be at least $s_0$ nodes of positive degree in the background graph. Assuming that outliers are selected uniformly at random, we expect $L + (n-L) (s_0 / n)$ nodes of positive degree in the background graph. (In fact, since there is a slight bias toward selecting small degree nodes for outliers and $L$ has a bias toward large degree nodes, we expect slightly more nodes of positive degree in the background graph, which is good.) We introduce the following constraint: a node of degree $w_i$ can become an outlier if
\begin{equation}\label{eq:outliers}
w_i \le \bar{L} + s_0 - \bar{L} s_0/n -1.
\end{equation}
Finally, $s_0$ nodes satisfying~(\ref{eq:outliers}) are selected uniformly at random to become outliers. (In the implementation, these nodes simply form an independent ``community'' with $y_i=0$ and $z_i=w_i$.) 

\subsubsection*{Assigning Nodes to Communities.}

Similarly to the potential problem with outliers, we need to make sure that non-outliers of a large degree are not assigned to small communities. Based on the parameter $\xi$ we know that roughly $(1-\xi) w_i$ neighbors of a node of degree $w_i$ will be present in its own community. However, this is only the lower bound as some neighbors in the background graph might end up there by chance. Hence, in order to make enough room in the community graph for all neighbors of a given node, the original \ABCD\ algorithm needs to compute $x_i$, the expected number of neighbors of a node of degree $w_i$ that end up in its own community. We need to recompute $x_i$ to incorporate the existence of outliers.

Assuming that nodes are assigned randomly with a distribution close to the uniform distribution, we expect $W s_0 / n$ points (in the corresponding configuration model) in the background graph to be associated with outliers, where $W := \sum_{i \in [n]} w_i$ is the volume of the graph (equivalently, the total number of points in the corresponding configuration model). Similarly, we expect $\xi$ fraction of the points associated with non-outliers to end up in the background graph, that is, $W (1-s_0/n) \xi$ points. In order to estimate what fraction of neighbors of a given non-outlier node is expected to be within the same community, we need to answer the following question: what is the probability that a random point in the background graph associated with a non-outlier is matched with a point within the same community? It is equal to
$$
\sum_{j \in [\ell]} \frac {s_j}{n-s_0} \cdot \frac { \frac {s_j}{n-s_0} W (1-s_0/n) \xi } { W (1-s_0/n) \xi +  W s_0 / n } 
= \sum_{j \in [\ell]} \left( \frac {s_j}{n-s_0} \right)^2 \frac {(n-s_0) \xi} {(n-s_0)\xi + s_0}.
$$
Indeed, with probability $\frac {s_j}{n-s_0}$ a random point belongs to community $j$. There are $\frac {s_j}{n-s_0} W (1-s_0/n) \xi$ points associated with community $j$ and the total number of points in the background graph is $W (1-s_0/n) \xi +  W s_0 / n$. Hence, one can easily estimate the probability that the point from community $j$ is matched with another point from the same community. The expected number of neighbors of a node of degree $w_i$ that stay within the same community is then
$$
x_i := \left( 1 - \xi + \xi \sum_{j \in [\ell]} \left( \frac {s_j}{n-s_0} \right)^2 \frac {(n-s_0) \xi} {(n-s_0)\xi + s_0} \right) w_i = (1-\xi \phi) w_i,
$$
where
$$
\phi := 1 - \sum_{j \in [\ell]} \left( \frac {s_j}{n-s_0} \right)^2 \frac {(n-s_0) \xi} {(n-s_0)\xi + s_0}.
$$
In particular, we expect $(1-\xi\phi)(1-s_0/n)$ fraction of edges to stay within one of the communities. Moreover, as expected, if $s_0=0$, then we recover the value of $\phi$ used in the original \ABCD\ model, namely,
$$
\phi = 1 - \sum_{j \in [\ell]} \left( \frac {s_j}{n} \right)^2.
$$
As in the original \ABCD\ model, a node of degree $w_i$ can be assigned to a community of size $s_j$ if $x_i \le s_j-1$. We select one admissible assignment of non-outliers to communities uniformly at random which turns out to be relatively easy from both theoretical and practical points of view.

\medskip

\begin{figure}
\centering
\includegraphics[scale=0.33]{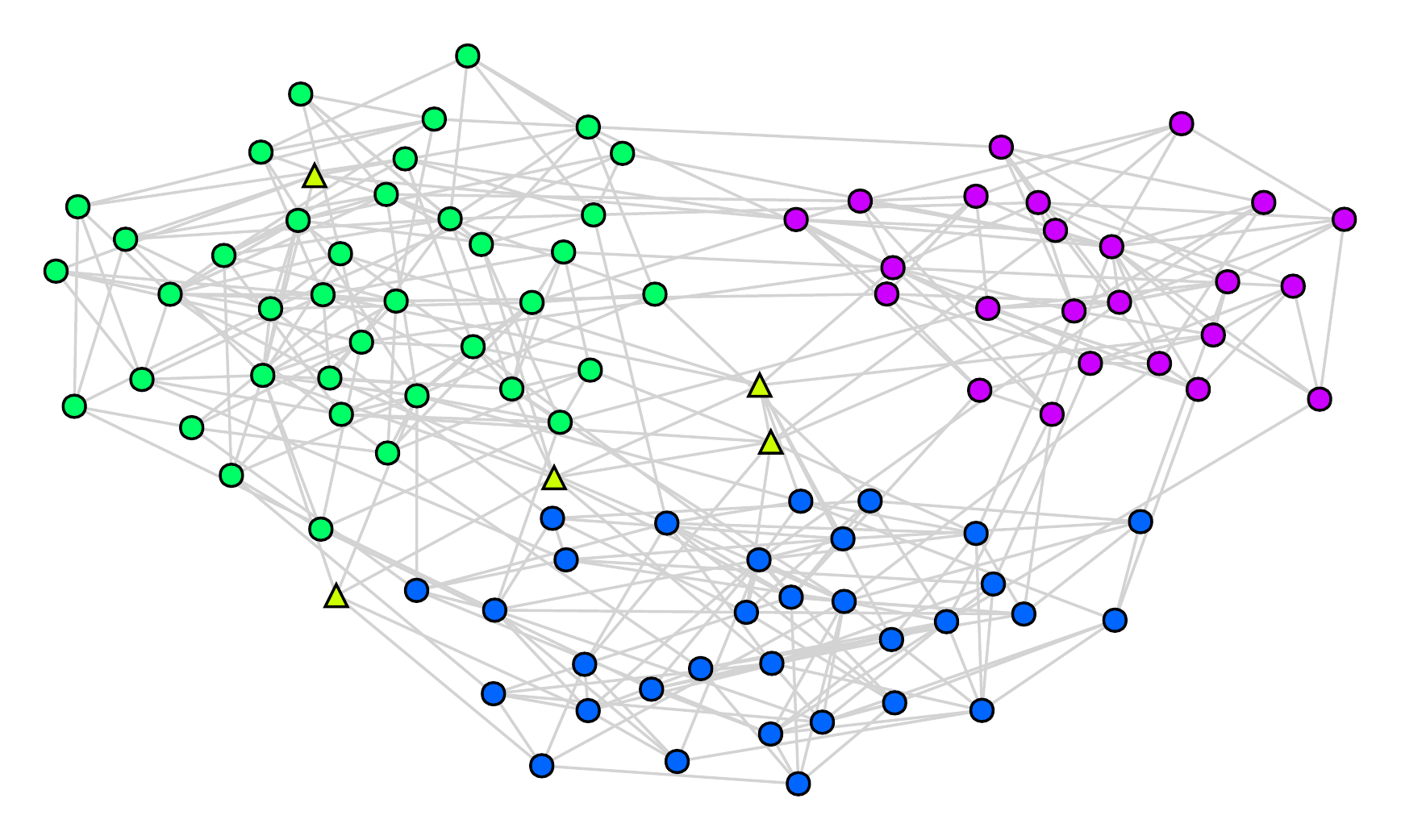}
\hspace{.1cm}
\includegraphics[scale=0.33]{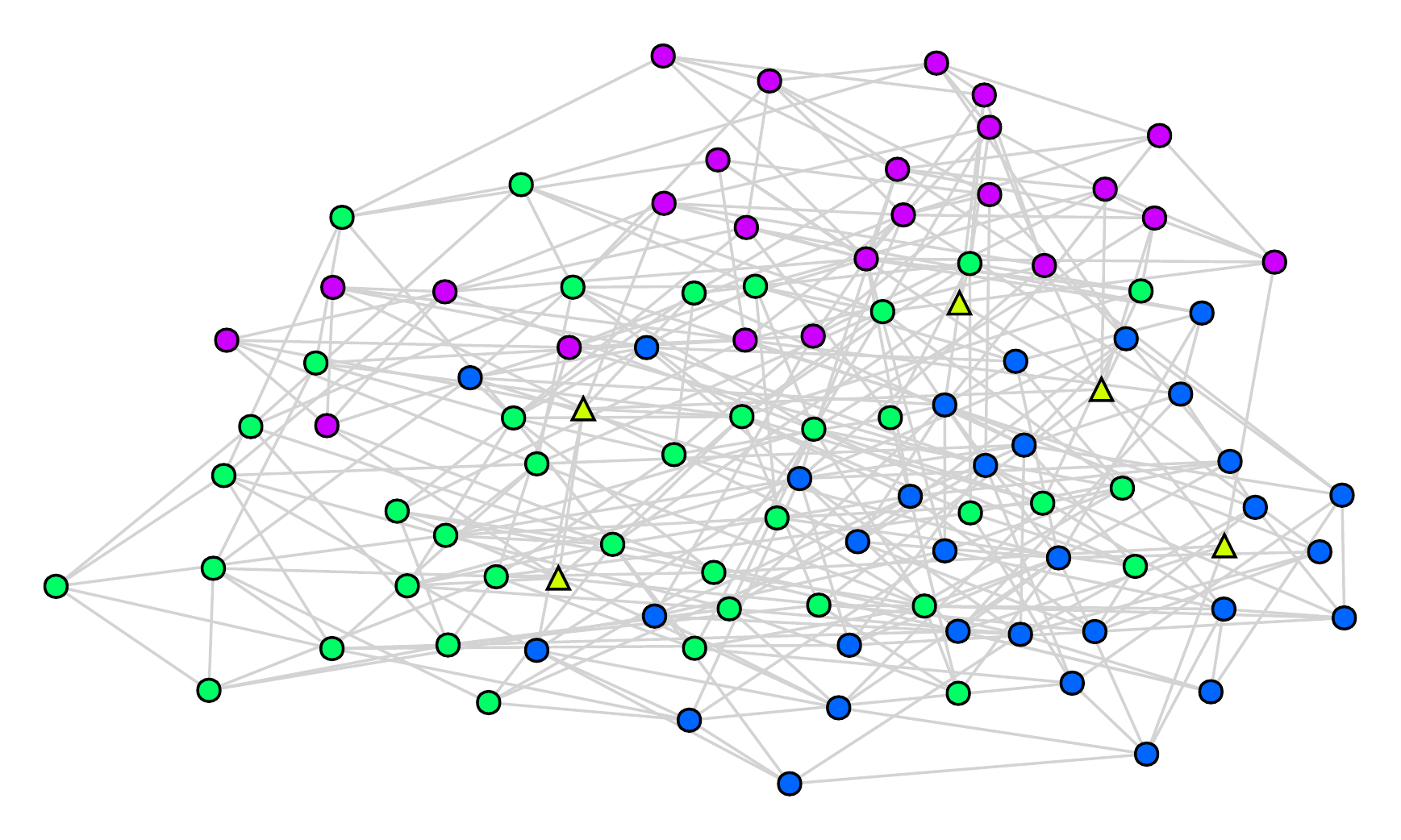}
\caption{Two examples of \textbf{ABCD+o} graphs with low level of noise ($\xi=0.2$, left) and high level of noise ($\xi=0.4$, right). The number of outliers is $s_0=5$ (depicted as triangles).}
\label{fig:examples+o}
\end{figure}

Two examples of \textbf{ABCD+o} graphs on $n=100$ nodes are presented in Figure~\ref{fig:examples+o} (as in the previous figure, a standard Fruchterman-Reingold layout was used to plot them). The number of outliers is $s_0=5$ and the remaining parameters are exactly the same as the ones to produce Figure~\ref{fig:examples}.  Communities are shown in different colors and outliers are displayed as triangles. The left plot has the mixing parameter set $\xi=0.2$ while the ``noisier'' graph on the right plot has the parameter fixed to $\xi=0.4$. In the left plot, it is visible that 4 out of 5 outliers are clearly located \emph{between} the communities, while one of them falls within a single community. (Recall that outlier nodes have all of their incident edges generated in the background graph. As a result, neighbors of outliers are selected randomly from the entire graph. For large networks, it will be highly unlikely that most neighbors belong to one community but for small graphs, it could happen with non-negligible probability which turned out to be the case in our experiment. After all, it is a stochastic process and natural fluctuations may and do occur.) In the right plot, which is noisier, we still see that outliers are surrounded by nodes belonging to different communities.

\subsection{Scalability}

The implementation of the algorithm is done in Julia language~\cite{bezanson2014}. It is an extension of the \textbf{ABCD} model implementation~\cite{kaminski2021artificial} and so it does not change its computational complexity. For this reason, as reported earlier in~\cite{kaminski2021artificial}, \textbf{ABCD+o} generates typical graphs on 10 million nodes in under 2 minutes, which is of the order of 100 times faster than the reference \textbf{LFR} algorithm implementation.

\section{Experiments---Distinguishing Properties of Outliers}\label{sec:experiments}

In order to better understand the properties of outliers, we perform a few experiments on the well-known College Football real-world network with known community structure and the presence of outliers. 
We consider four different ways to detect outliers.
In order to show that our new \textbf{ABCD+o} model exhibits similar desired properties, we generated graphs on $n=10{,}000$ nodes with $s_0=500$ outliers (5\%). Degree distribution was generated with power law exponent $\gamma=2.5$ with minimum and maximum values of 5 and 500, respectively. Community sizes were generated with power law exponent $\beta = 1.5$ with minimum and maximum values 100 and 1{,}000, respectively. We independently generated graphs for all values of $\xi \in \{0.1, 0.2, \ldots, 1.0\}$ but the degree distribution and the distribution of community sizes were coupled 
so that all 10 graphs use the same distributions.

With the experiments presented in this section, we illustrate the usefulness of benchmarks such as \textbf{ABCD+o} to compare various anomaly detection methods under different scenarios such as graphs with more or less noise, nodes with varying degrees, etc.

\subsection{The College Football Graph}\label{sec:football}

\begin{figure}
\centering
\includegraphics[scale=0.5]{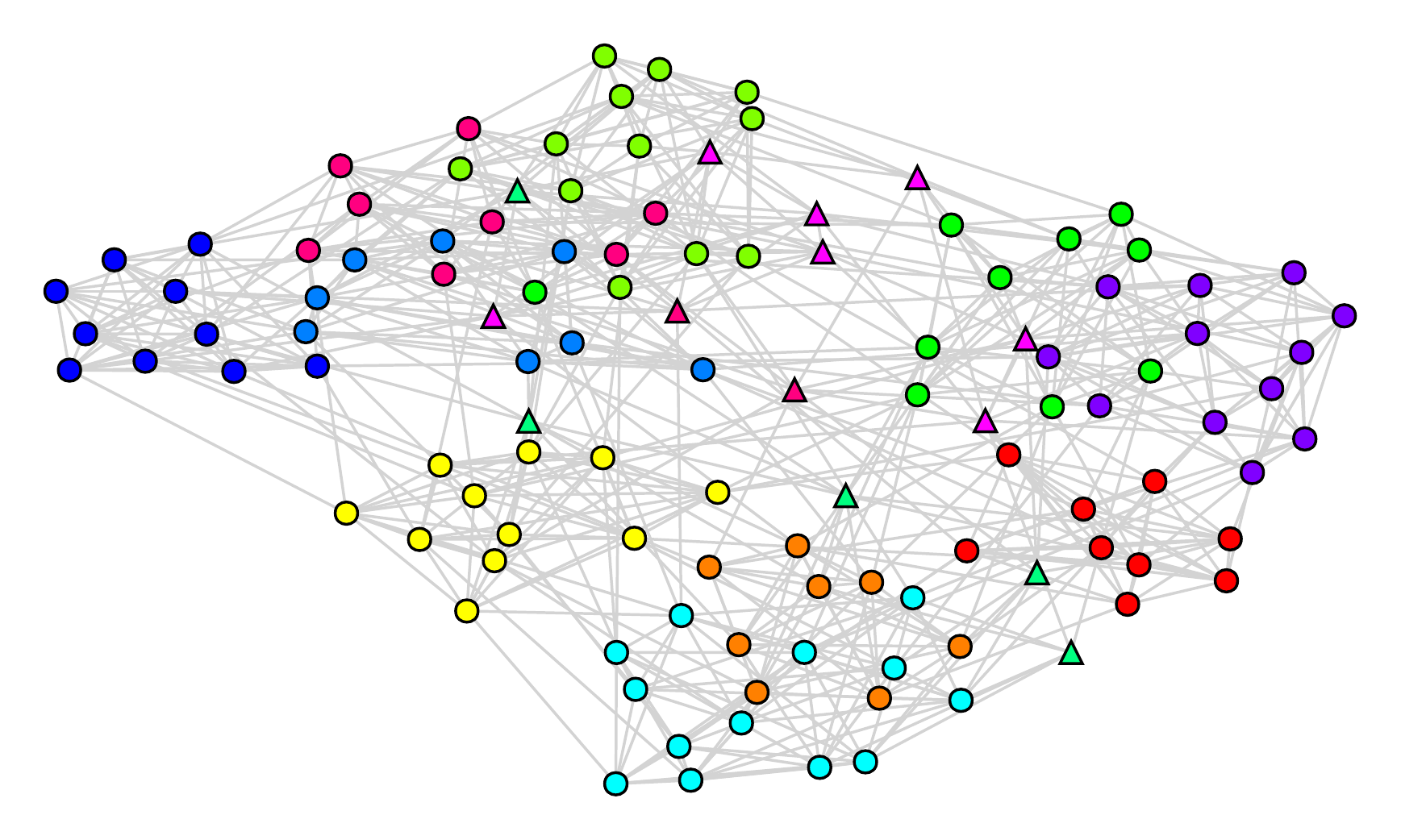}
\caption{The College Football Graph; outliers are displayed triangles.}
\label{fig:football}
\end{figure}

The College Football real-world network represents the schedule of United States football games between Division IA colleges during the regular season in Fall 2000~\cite{girvan2002community}. The data consists of 115 teams (nodes) and 613 games (edges). The teams are divided into conferences containing 8--12 teams each. In general, games are more frequent between members of the same conference than between members of different conferences, with teams playing an average of about seven intra-conference games and four inter-conference games in the 2000 season. There are a few exceptions to this rule, as detailed in~\cite{lu2018community}: one of the conferences is really a group of independent teams, one conference is really broken into two groups, and 3 other teams play mainly against teams from other conferences. As it is commonly done in the literature, we refer to those 14 teams as outlying nodes, which we represent with distinctive triangles in Figure~\ref{fig:football}.

\subsection{Participation Coefficient}\label{sec:participation}

The following definitions are commonly used in the literature~\cite{flake2000efficient,radicchi2004defining} (see also~\cite{kaminski2021mining}). We say that a set of nodes $C \subseteq V$ forms a \emph{strong community} if each node in $C$ has more neighbors in $C$ than outside of $C$. One may relax this strong notion and say that $C$ forms a \emph{weak community} if the average degree inside the community $C$ (over all nodes in $C$) is larger than the corresponding average number of neighbors outside of $C$. In this context, an \emph{outlier} could be formally defined as a node that does not have the majority of its neighbors in any of the communities. In the \textbf{ABCD+o} model, non-outliers are expected to have more than half of their neighbors in their own community, provided that $\xi < 0.5$. On the other hand, outliers in our model are expected to satisfy the above-desired property, unless there is an enormous community spanning more than 50\% of nodes. 

A more refined picture is provided by the next coefficient which is a natural measure of concentration. For any partition $\A = \{A_1, \ldots, A_\ell\}$ of the set of nodes, the \emph{participation coefficient} of a node $v$ (with respect to $\A$) is defined as follows:
$$
p(v) = 1 - \sum_{i=1}^{\ell} \left( \frac {\deg_{A_i}(v)}{\deg(v)} \right)^2,
$$
where $\deg_{A_i}(v)$ is the number of neighbours of $v$ in $A_i$. The participation coefficient $p(v)$ is equal to zero if $v$ has neighbors exclusively in one part. Members of strong communities satisfy, by definition, $p(v) < 3/4$. In the other extreme case, the neighbors of $v$ are homogeneously distributed among all parts and so $p(v)$ is close to the trivial upper bound of 
$$
1 - \sum_{i=1}^{\ell} \left( \frac {\deg(v)/\ell}{\deg(v)} \right)^2 = 1-\frac {1}{\ell} \approx 1.
$$

\begin{figure}
\centering
\includegraphics[scale=0.38]{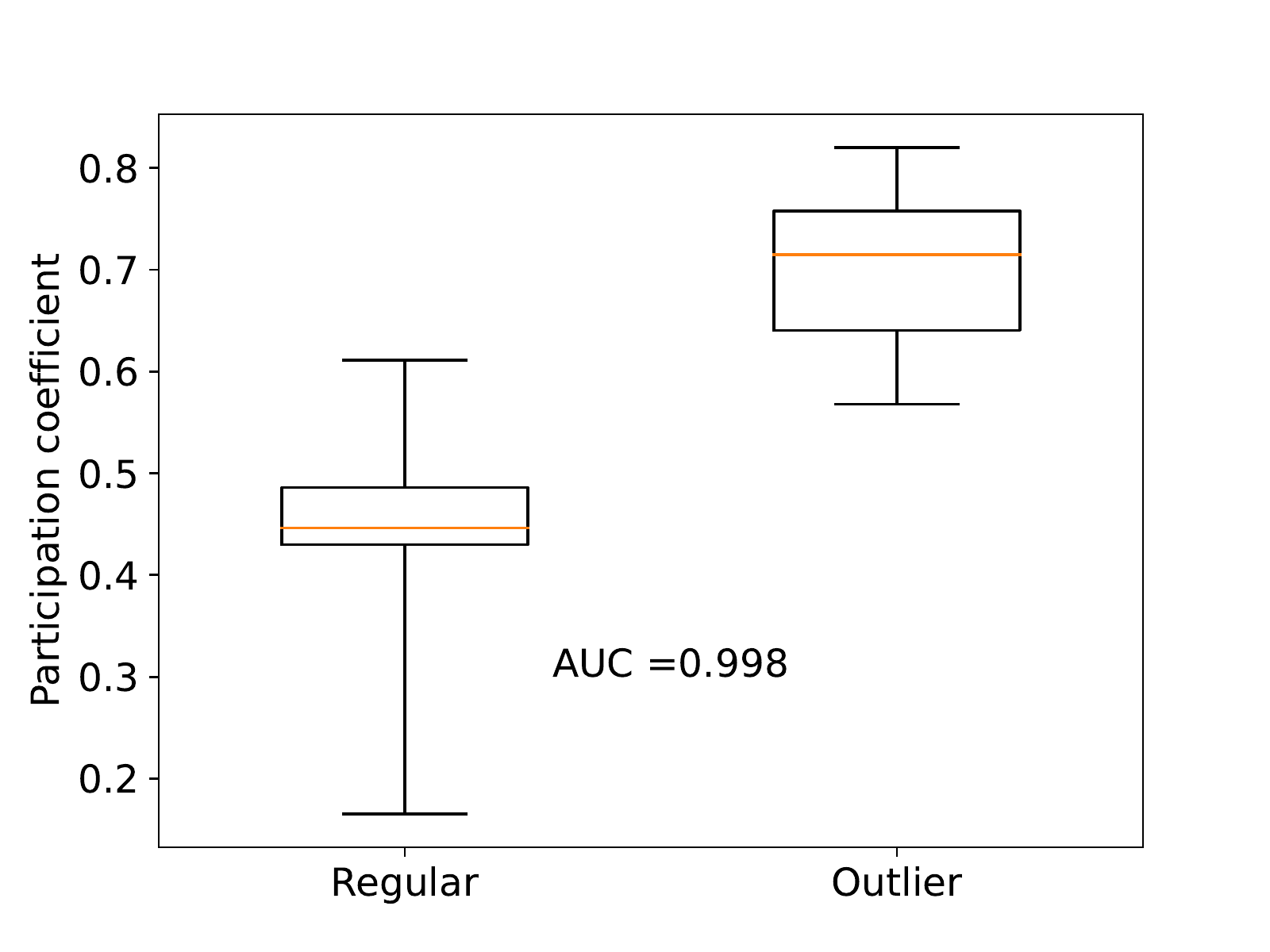} 
\hspace{-.5cm}
\includegraphics[scale=0.38]{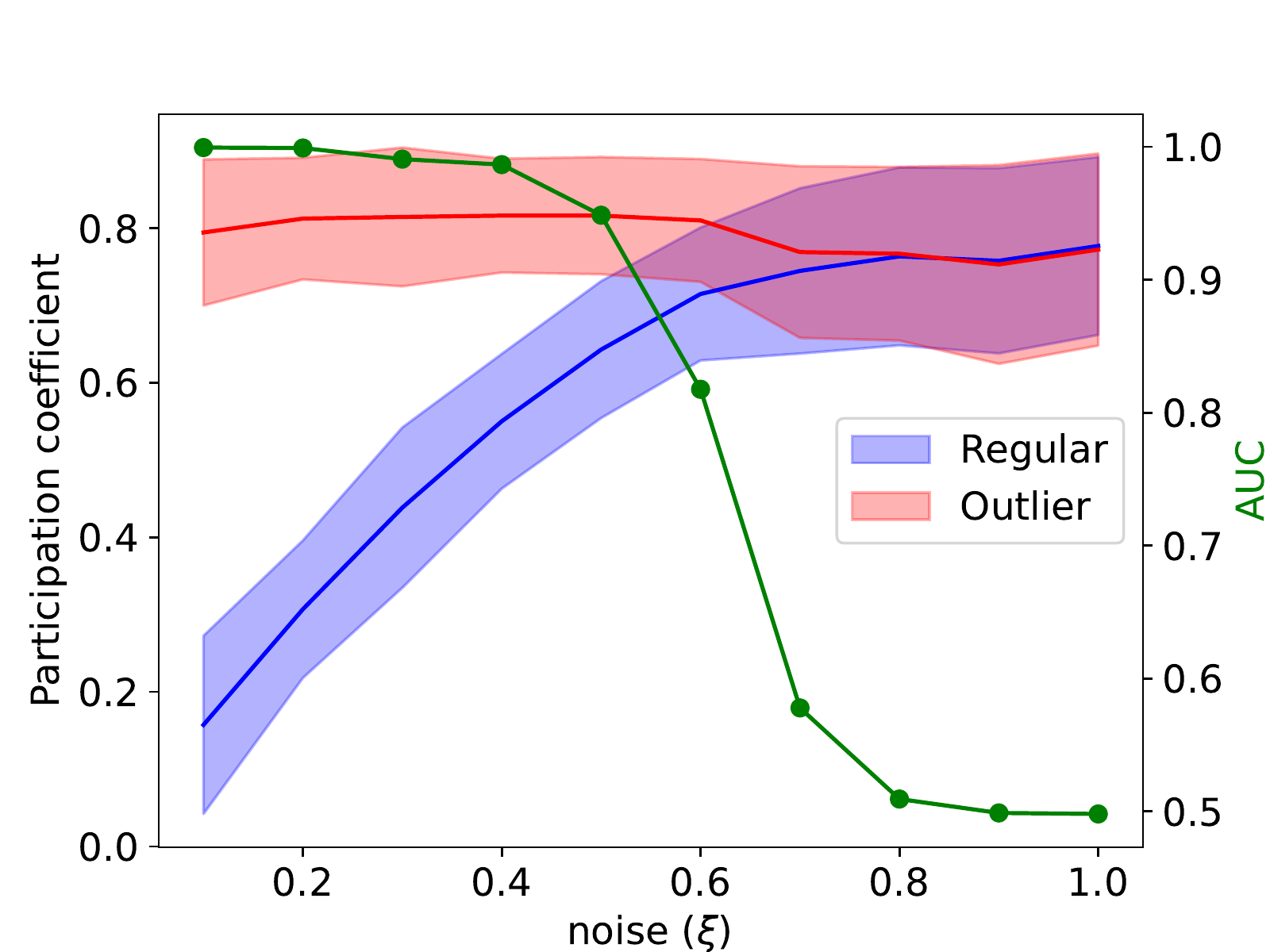}
\caption{Distribution of the \emph{participation coefficient} for regular and outlier nodes: College Football Graph (left) and \textbf{ABCD+o} model (right).}
\label{fig:participation}
\end{figure}

For the experiments shown below, even though we have the ground truth communities available to use (which is almost always not the case in practice), we computed the participation coefficients using communities (partition $\A$) we obtained with the \textbf{ECG} clustering algorithm (we describe this algorithm in the following subsection). The distribution of the participation coefficient among outliers and non-outliers for the College Football Graph is presented on the box plot in Figure~\ref{fig:participation} (left). We see that outliers have a significantly larger average value of $p(v)$ than the corresponding value for non-outliers. 
We also computed the AUC score (the area under the ROC curve), which corresponds to the probability that a randomly selected outlier node has a larger score than a randomly selected non-outlier node. We see that this value is almost one (approximately $0.998$), showing a very good separation between the two classes.

The corresponding averages (together with associated standard deviations) for the \textbf{ABCD+o} model with different levels of noise are presented in Figure~\ref{fig:participation} (right). For a low level of noise (small values of $\xi$) there is a clear difference between outliers and non-outliers but the discrepancy diminishes for noisy graphs (large values of $\xi$). In the extreme case when $\xi=1$ there is no difference between the two classes and so the averages are close to each other as they should be.
This is also well illustrated by the corresponding AUC scores we computed for various values of $\xi$, showing very good class separation for small to mid-range values of $\xi$, but tending toward 0.5 for large $\xi$, which amounts to having no separation between the two classes.

\subsection{ECG Coefficient}\label{sec:ECG}

\textbf{Ensemble Clustering algorithm for Graphs} (\textbf{ECG})~\cite{poulin2018ensemble}\footnote{\url{https://github.com/ftheberge/graph-partition-and-measures}} is a consensus clustering method based on the classical \textbf{Louvain} algorithm. A convenient ``side-effect'' of this algorithm, which is useful from the perspective of our experiments, is that it can be used to define simple scores to identify outliers. In its first phase, several low-level partitions are computed with different randomization (the ensemble). In the next phase, for each edge, we compute the proportion of partitions in the ensemble where both nodes incident to this edge were assigned to the same community. Those are the \emph{ECG edge scores}. High scores are indicative of stable pairs that often appear in the same part. For a given node $v$, we define $E(v)$ to be the average ECG score over all edges incident to $v$, and we call it the \emph{ECG coefficient} of a node $v$.  It is expected that outliers are more challenging to cluster which should be manifested by relatively small ECG coefficients $E(v)$ associated with these nodes.

\begin{figure}
\centering
\includegraphics[scale=0.38]{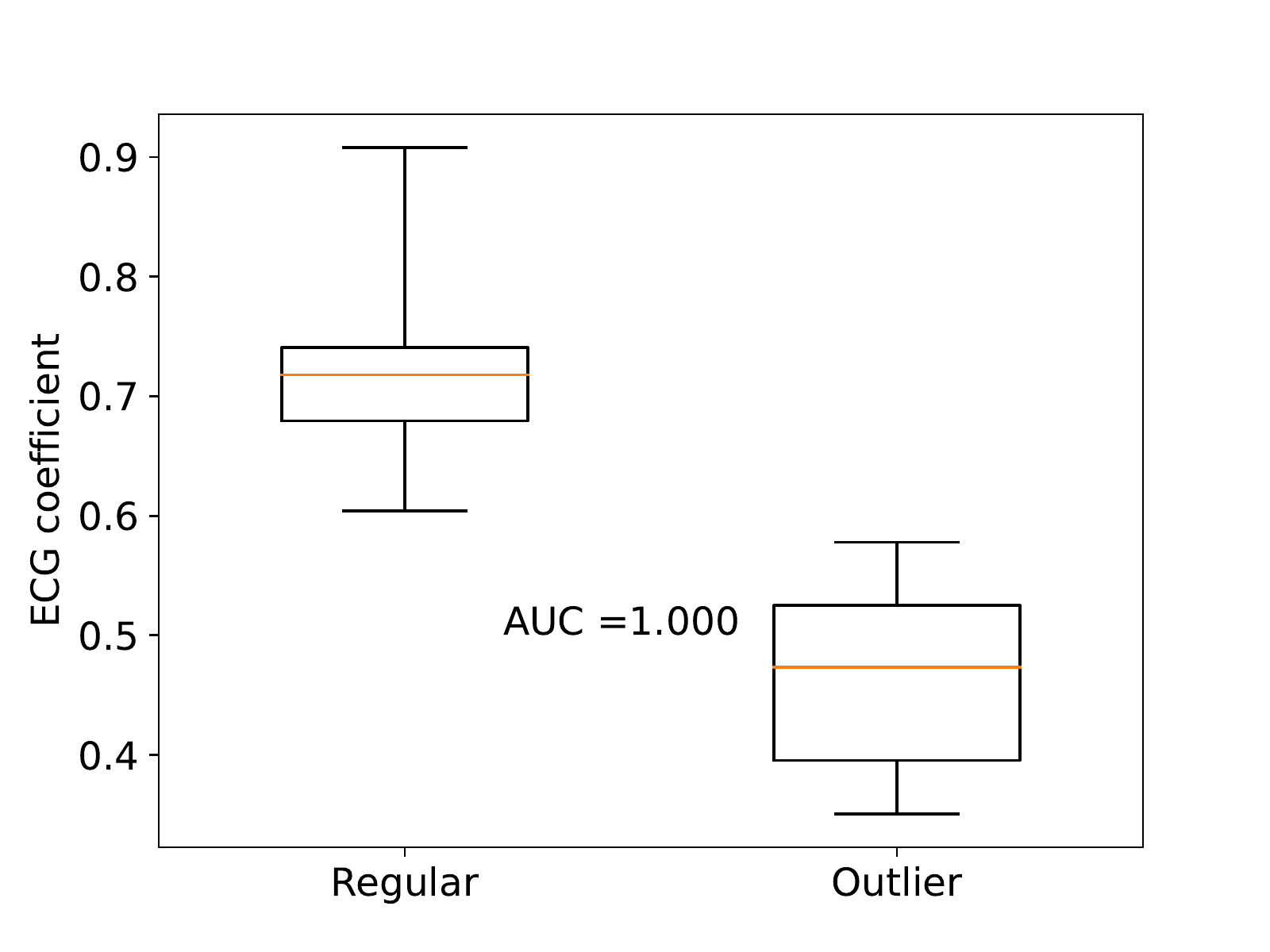}
\hspace{-.5cm}
\includegraphics[scale=0.38]{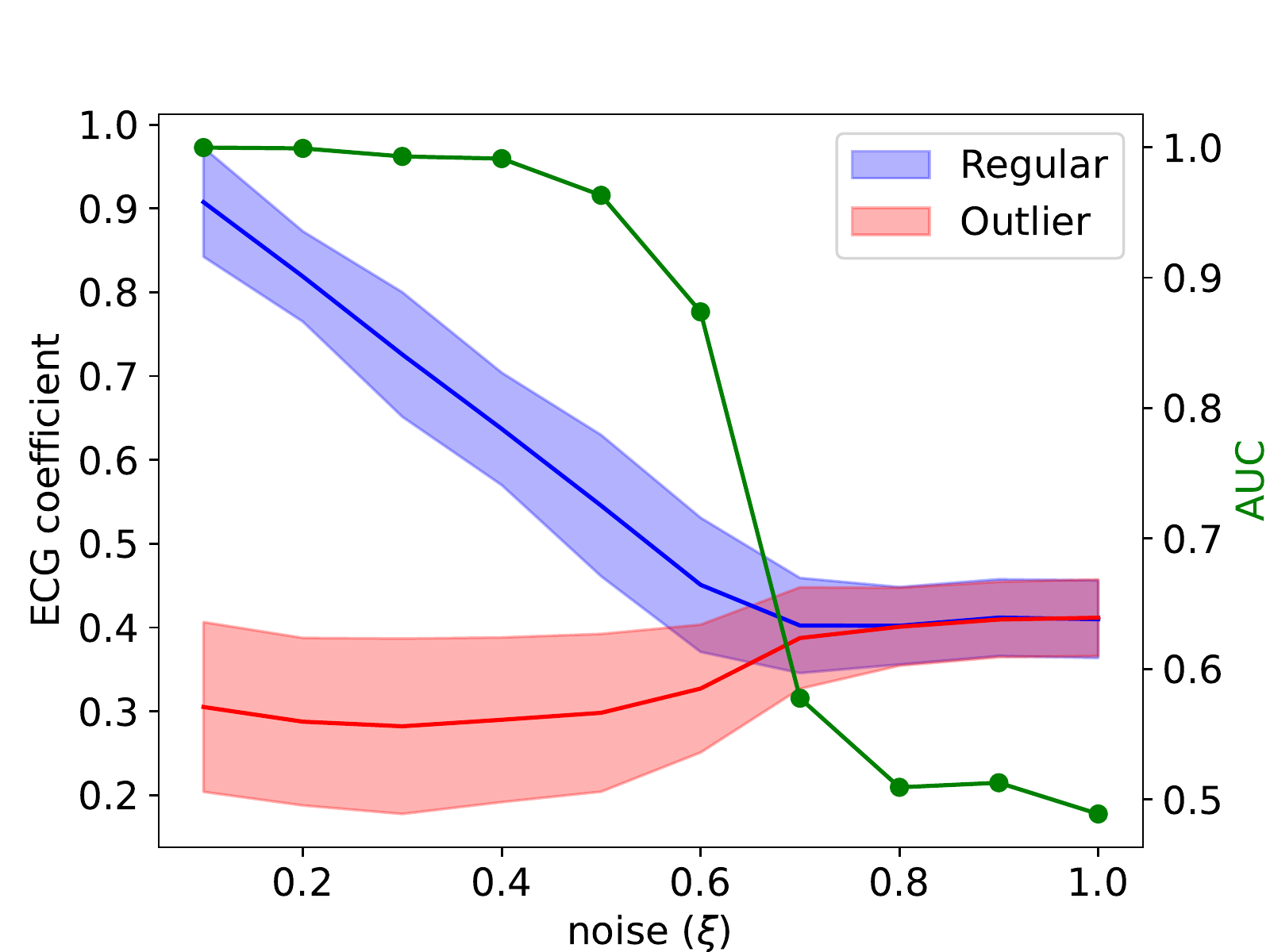}
\caption{Distribution of the \emph{ECG coefficient} for regular and outlier nodes: College Football Graph (left) and \textbf{ABCD+o} model (right).}
\label{fig:ecg}
\end{figure}

As it was done for the participation coefficient, we investigate the distribution of the ECG coefficient among outliers and non-outliers for the College Football Graph---see Figure~\ref{fig:ecg} (left). We see that it is another distinguishing coefficient---outliers have a significantly smaller average value of $E(v)$ than the corresponding value for non-outliers, and the AUC shows perfect separation between the classes.
Similar conclusions can be derived from the corresponding averages for the \textbf{ABCD+o} model---see Figure~\ref{fig:ecg} (right). As before, the difference becomes less visible as more noise is present. 

\subsection{Community Association Strength}\label{sec:edges}

With the participation coefficient we described earlier, we consider the distribution of communities amongst each node's neighbors, so that nodes that are strongly associated with one of the communities have skewed distributions. Here, given some node $v$, we only consider the node's own community $A_i$ and compute the fraction of edges that are within this community, namely, $\deg_{A_i}(v)/\deg(v)$. We then subtract the expected number of such edges under random assignment (approximately $\vol(A_i) / \vol(V)$) to obtain each node's \emph{community association strength}: 
$$
d(v) = \frac{\deg_{A_i}(v)}{\deg(v)} - \frac{\vol(A_i)}{\vol(V)}.
$$

\begin{figure}
\centering
\includegraphics[scale=0.38]{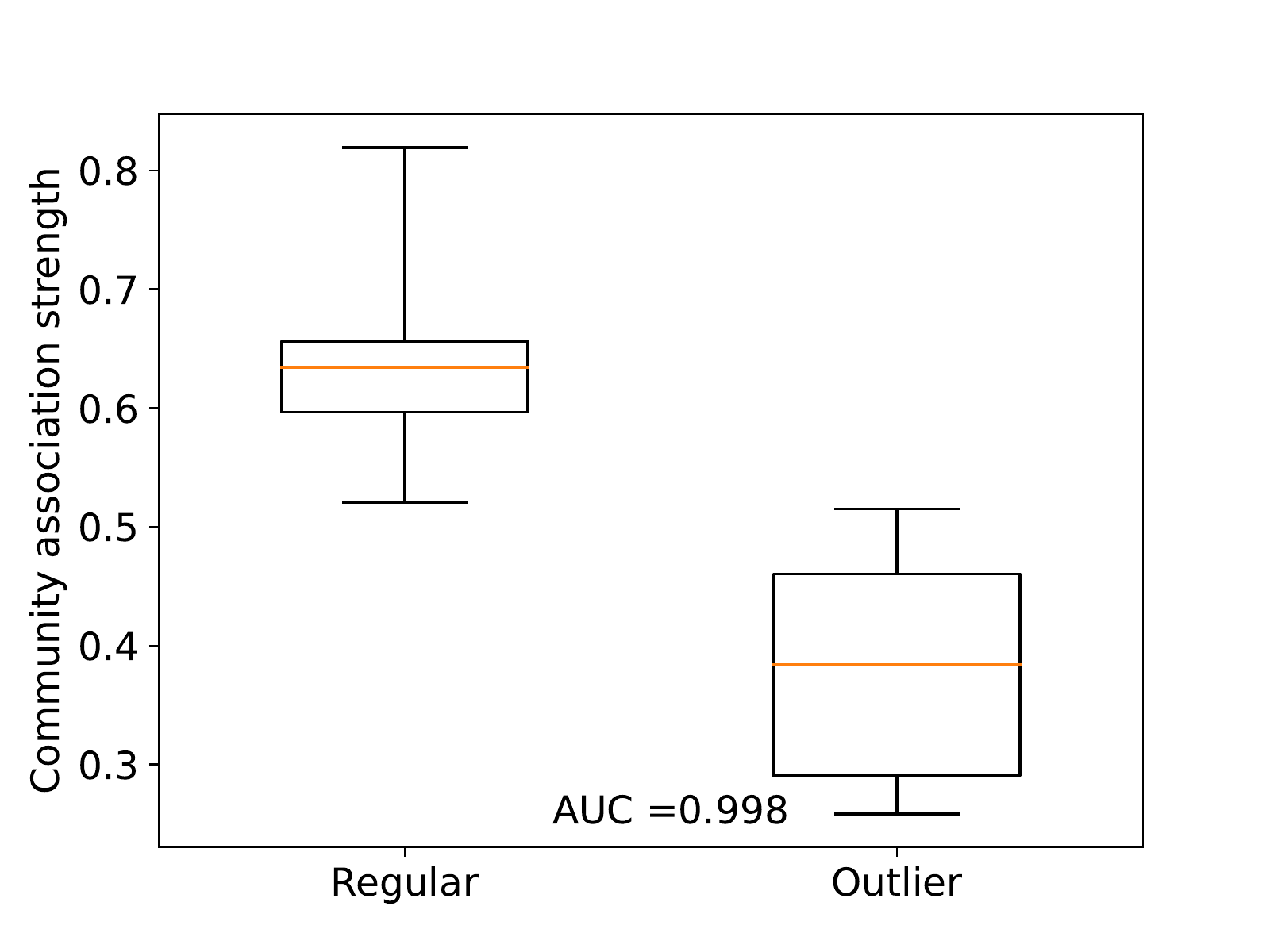}
\hspace{-.5cm}
\includegraphics[scale=0.38]{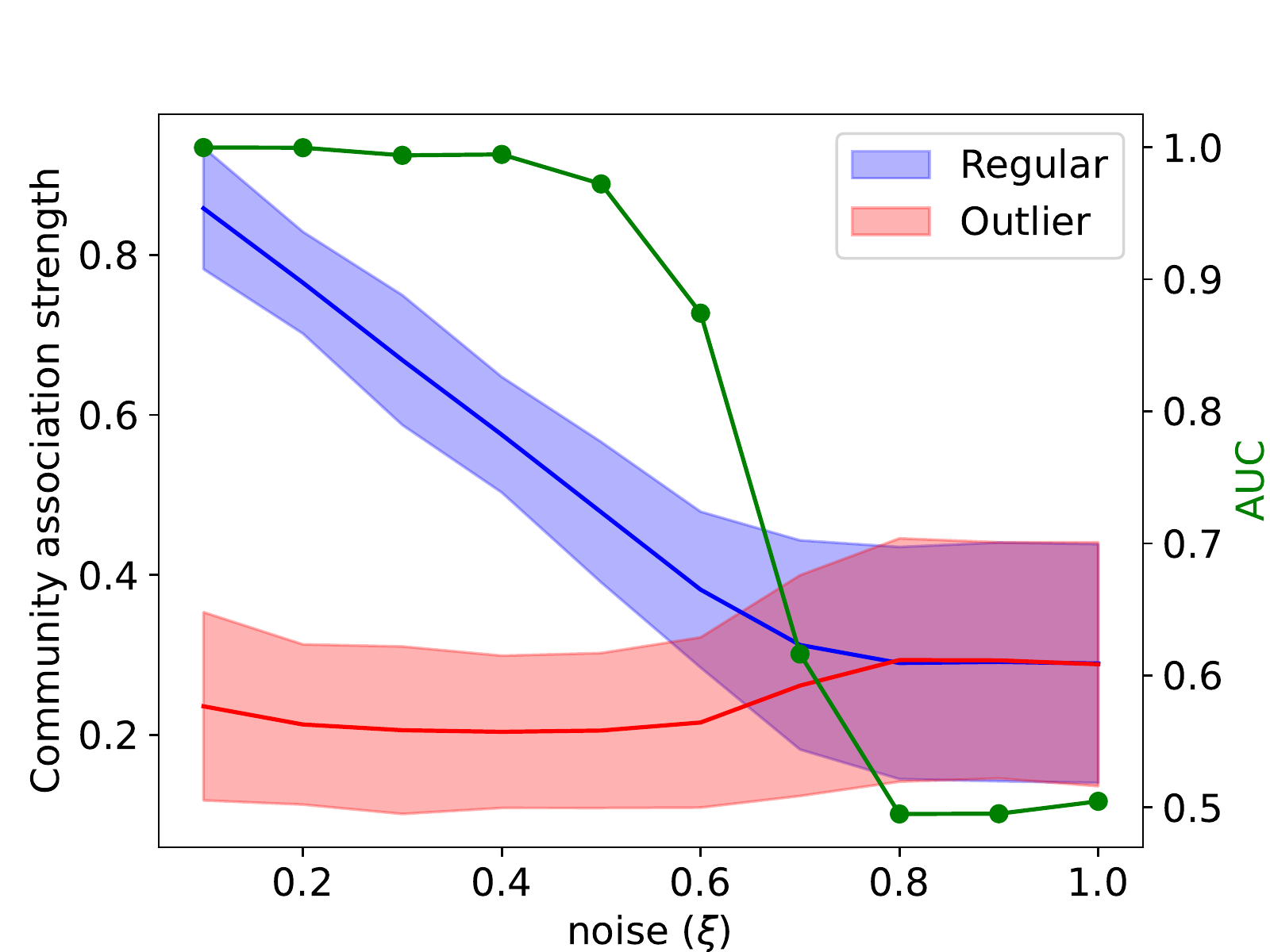}
\caption{Distribution of the \emph{community association strength} for regular and outlier nodes: College Football Graph (left) and \textbf{ABCD+o} model (right).}
\label{fig:diff}
\end{figure}

We repeat the same experiments as we did for the previous methods. Results are shown in Figure~\ref{fig:diff}, with very similar results and conclusions as with the previous two methods.

\subsection{Impact of Node Degree}
\label{sec:degree}

In this subsection, we reconsider the three methods presented so far but this time we group the nodes by their degrees for the \textbf{ABCD+o} graphs. We distinguish three families of nodes with respect to their degrees: (i) low-degree nodes (of degree 7 or less), (ii) medium-size degree nodes (from 8 to 20) and (iii) high-degree nodes (over 20). With this split, low-degree nodes make up over 50\% of the nodes, medium size a little under 40\%, and high-degree nodes about 10\%. For the College Football Graph introduced earlier, degree distribution is very homogeneous, so we do not consider it in this analysis.

\begin{figure}
\centering
\includegraphics[scale=0.36]{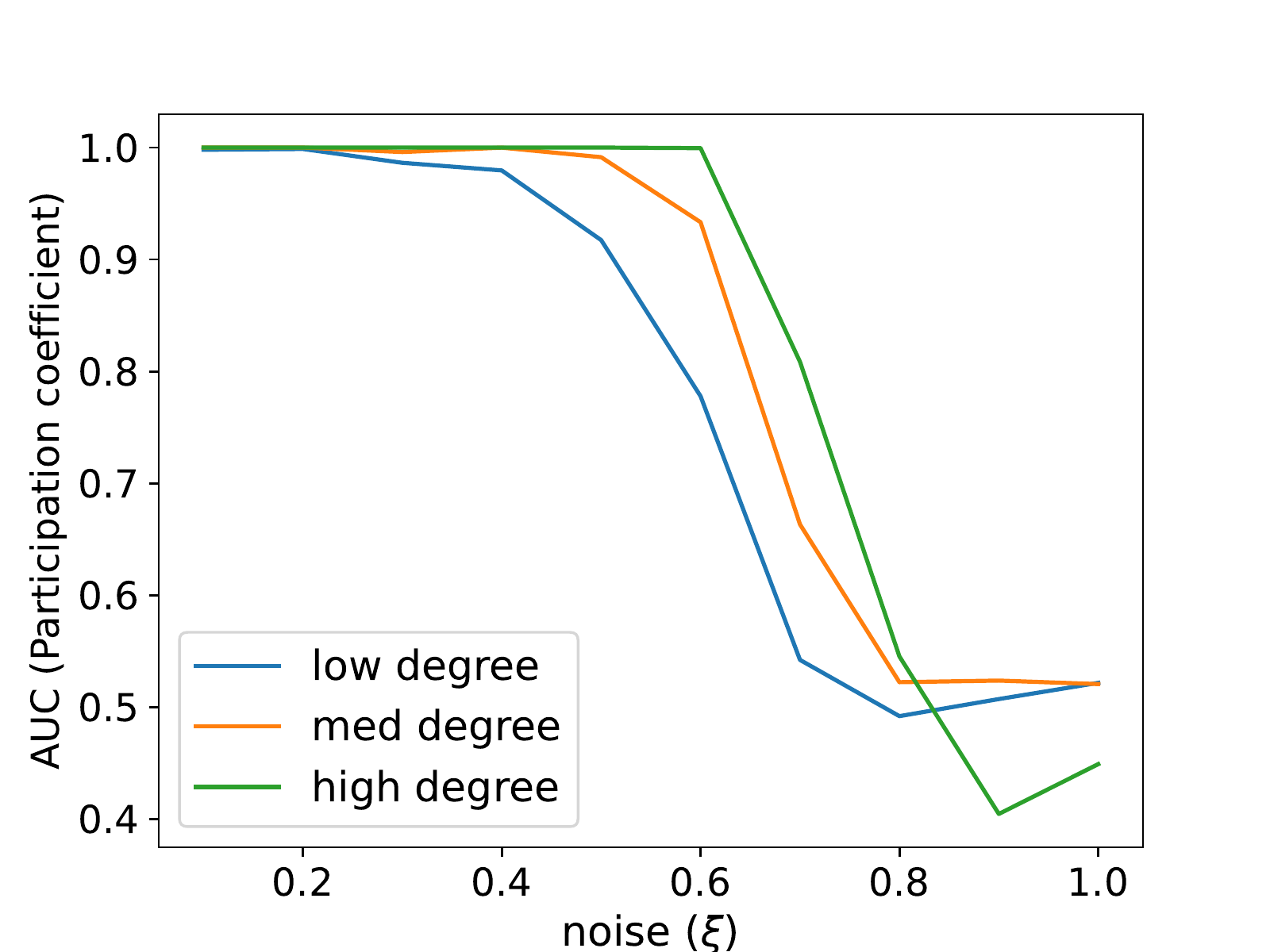}
\hspace{-.5cm}
\includegraphics[scale=0.36]{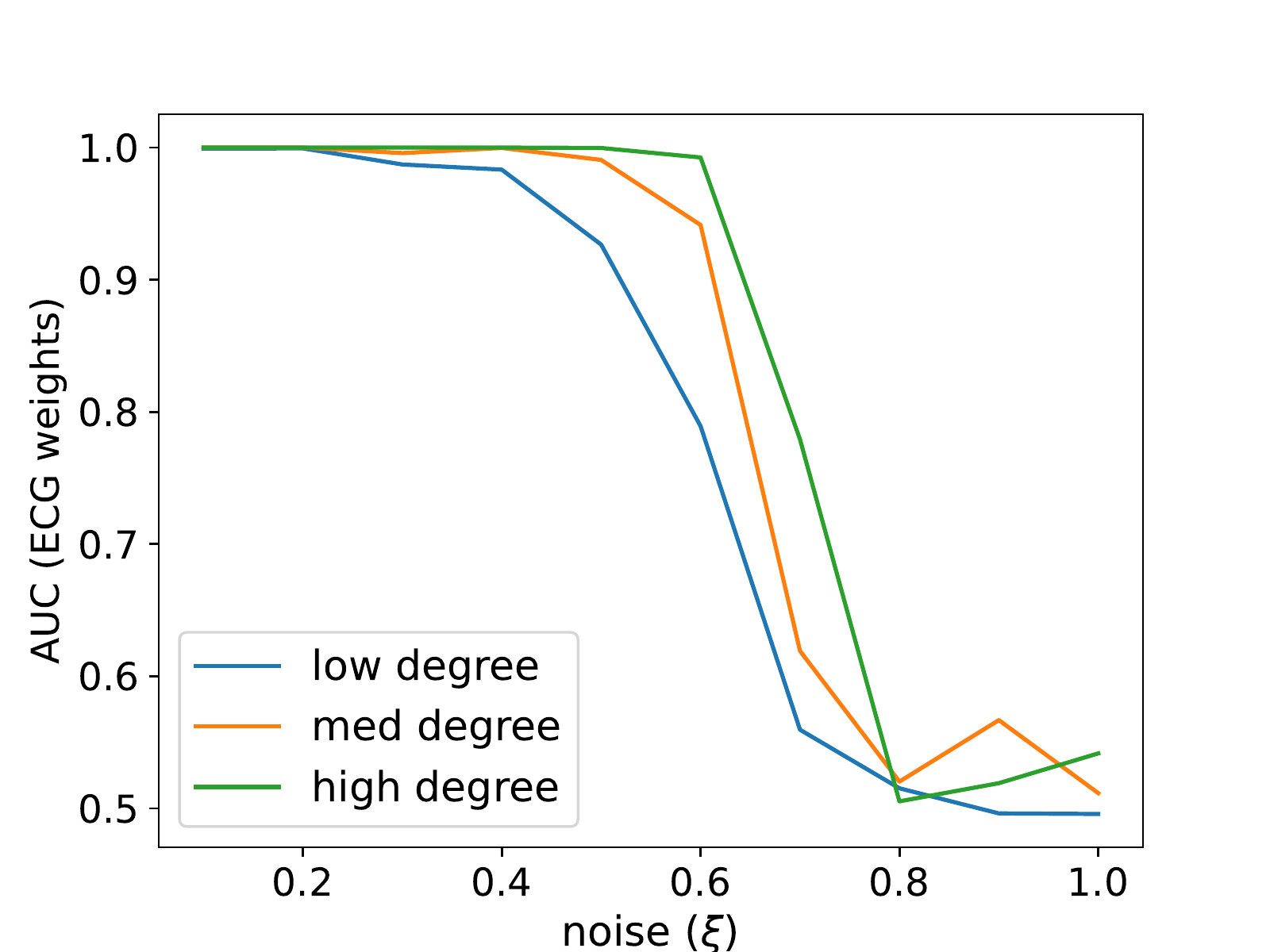}
\hspace{-.5cm}
\includegraphics[scale=0.36]{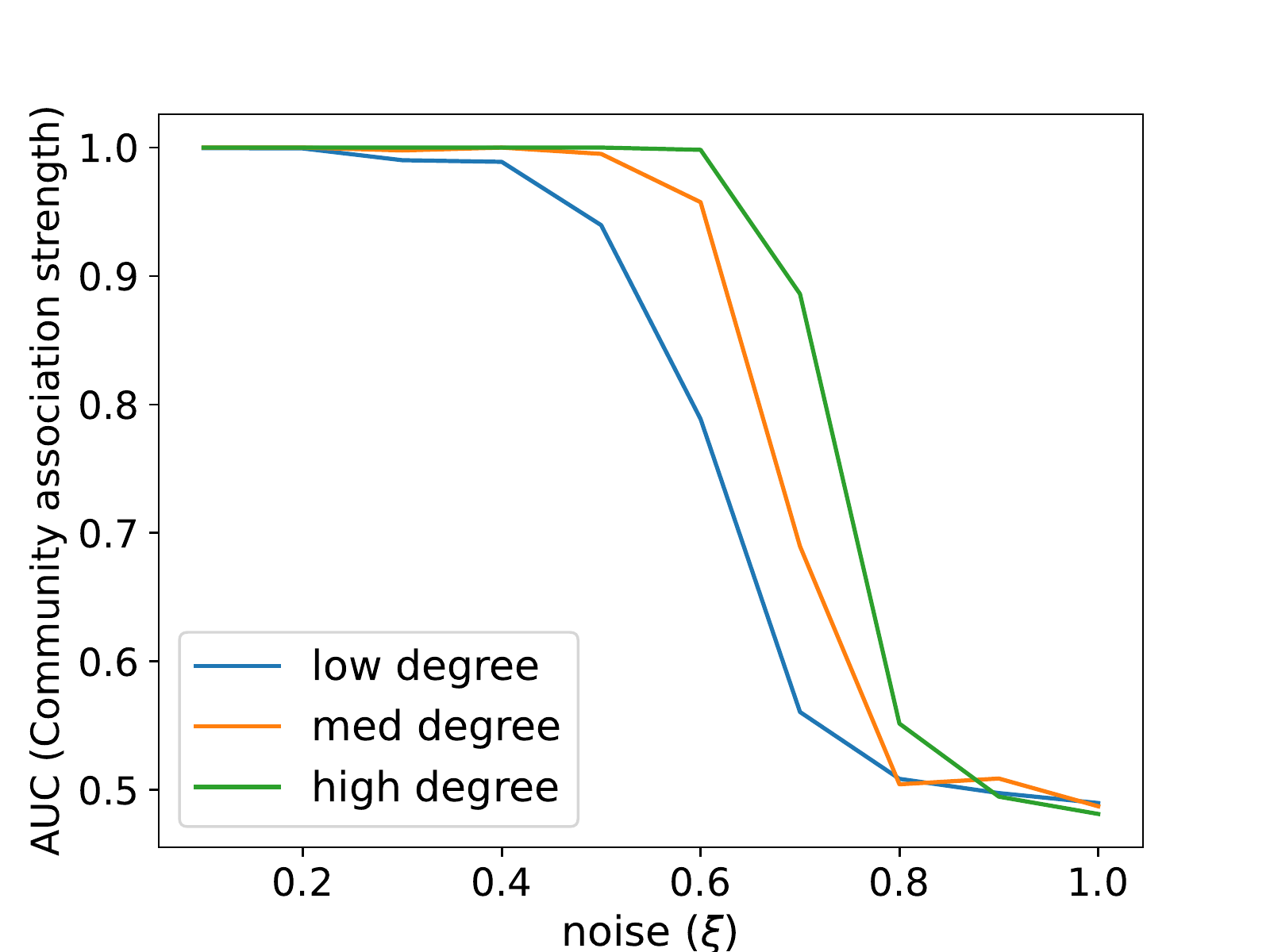}
\caption{AUC scores grouped by node degrees for the \textbf{ABCD+o} model.}
\label{fig:degree}
\end{figure}

Let us note an important property that in the \textbf{ABCD+o} model, the degree itself is not a discriminating feature between outlier and regular nodes, as both types follow the same expected degree distribution. For the graphs we generated, we have 5.1\% of outliers for the low and medium-size degree nodes, and 4.3\% for the (less frequent) high-degree nodes. In Figure~\ref{fig:degree}, we show the AUC scores for each method and node degree category in the \textbf{ABCD+o} graphs. It can be seen that it is slightly easier to detect high-degree outliers.

\subsection{Entropy---Geometric Chung-Lu Model}\label{sec:entropy}

For this experiment, we use node embeddings that recently gain a lot of interest. For each graph considered, we ran the \textbf{node2vec} embedding algorithm~\cite{grover2016node2vec} over a range of parameters, and we selected the best one using an ``unsupervised framework for comparing graph embeddings''~\cite{kaminski2020unsupervised,kaminski2021multi}\footnote{\url{https://github.com/KrainskiL/CGE.jl}}. This framework is based on a {\em geometric Chung-Lu} model, which allows the computation of edge probability in embedded space. With such selected embedding at hand, for each node $v$, one can compute $p_{v,i}$, the expected fraction of neighbors of $v$ that are in the community $i \in [\ell]$, assuming that there are $\ell$ communities found by some algorithm (we used ECG). From this distribution, we compute the \emph{entropy} for each node in the network: $H(v) = - \sum_{i \in [\ell]} p_{v,i} \ln(p_{v,i})$. High entropy is an indicator of anomalies so we can use it to rank the nodes from the most likely to the least likely to be anomalous. 

\begin{figure}
\centering
\includegraphics[scale=0.38]{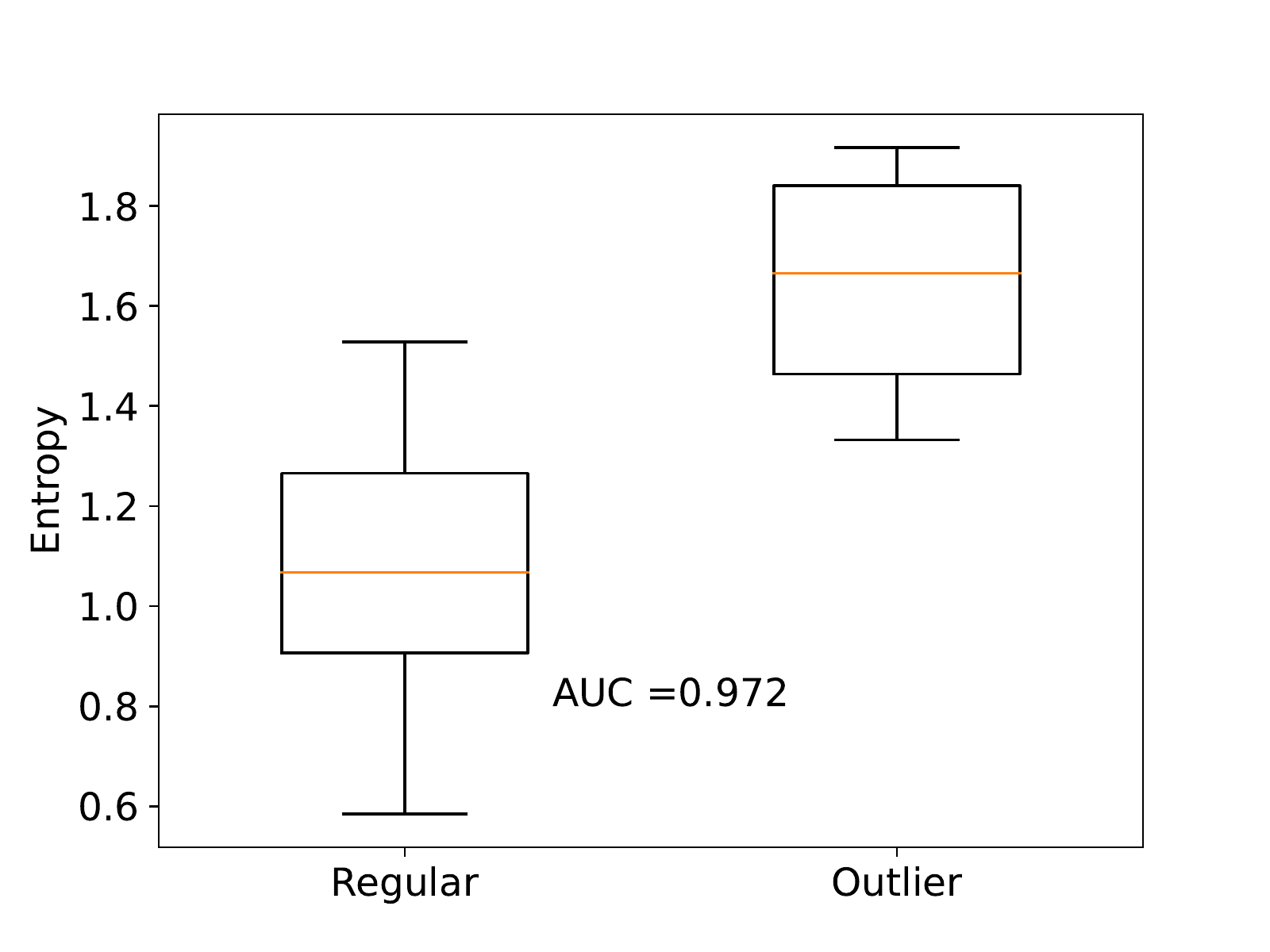}
\hspace{-.5cm}
\includegraphics[scale=0.38]{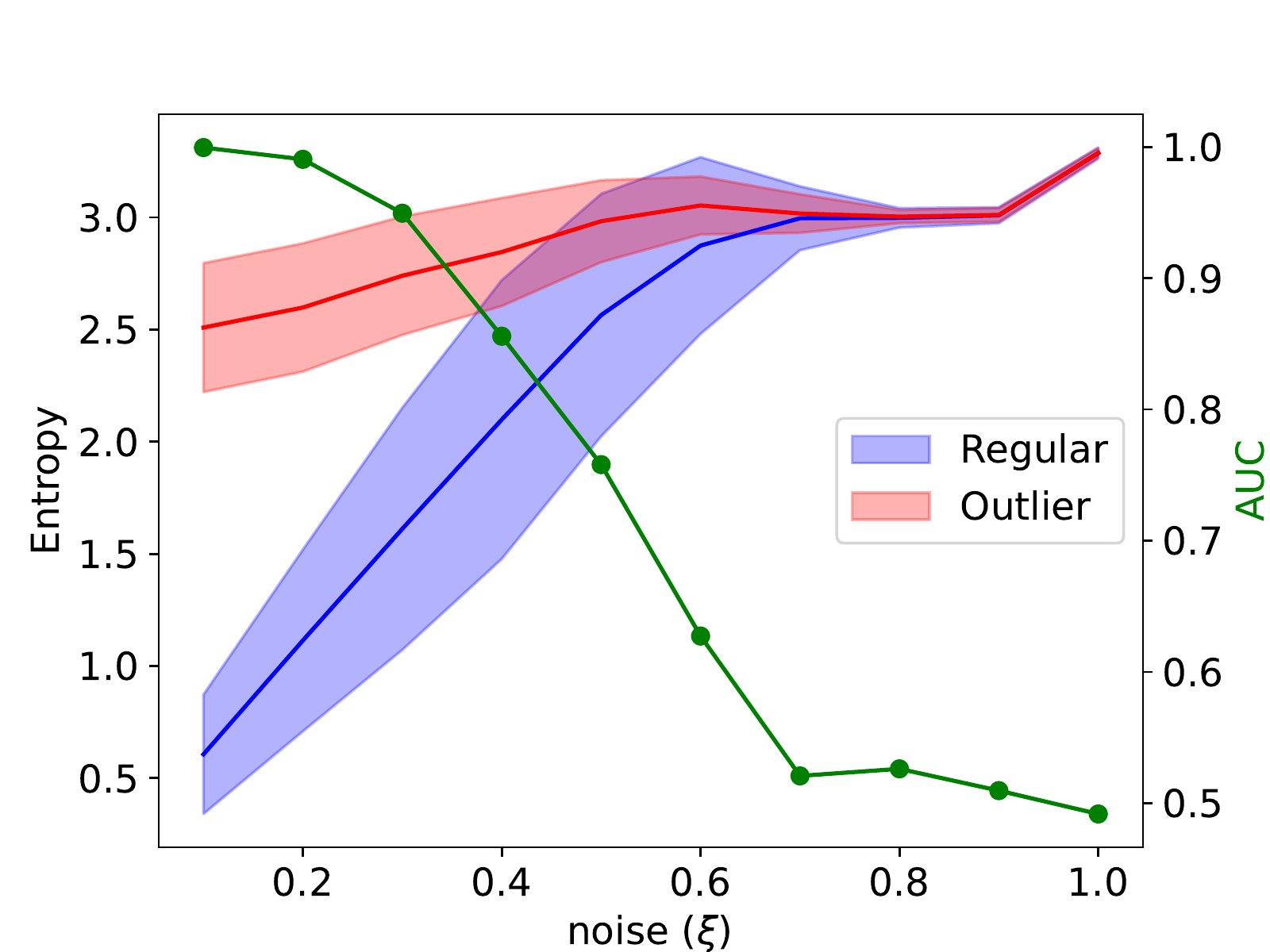}

\caption{Distribution of the \emph{entropy} for regular and outlier nodes: College Football Graph (left) and \textbf{ABCD+o} model (right).}
\label{fig:entropy}
\end{figure}

Results of the set of experiments for the College Football Graph and for several \textbf{ABCD+o} graphs are shown in Figure~\ref{fig:entropy}.
In both cases, while some good class separation can be observed, the separation is not as strong as with the three methods introduced earlier.

\subsection{Node Properties}\label{sec:properties}

In this subsection, we investigate the possible use of various node centrality measures as a way to distinguish regular nodes from outliers.
For the \textbf{ABCD+o} model, we grouped the nodes into three families with respect to their degrees, as we did in earlier experiments, and compared distributions of four different centrality measures: closeness centrality, eigen-centrality, PageRank, and betweenness. All plots are provided in the Appendix (see Figures~\ref{fig:close}--\ref{fig:between}). We also plot those four centrality measures for the College Football Graph, where the proportion of ``noise'' edges is about 0.37 (see Figure~\ref{fig:foot_features}).

From those experiments, we see a slight difference in the distribution of closeness centrality (Figure~\ref{fig:close}) and betweenness (Figure~\ref{fig:between}) for low-noise graphs. For closeness centrality, this can be explained by the fact that with low noise, non-outlier nodes have most of their edges within their community, thus are not very central in that sense. Outlier nodes have a higher betweenness since they act as bridges between the communities.
In the case of the College Football Graph, we also do not see much discriminative power in the distributions except for a slight difference in the betweenness scores (see Figure~\ref{fig:foot_features}).
In general, except for graphs with a very low noise level (that is, almost pure communities), it seems that such measures are not enough to distinguish regular (community) nodes and outliers.
This indicates that, indeed, specialized methods for community outlier detection are needed and that \textbf{ABCD+o} model has similar properties to real-world networks.

\subsection{Real Graph Example}
\label{sec:emailEu}

In this section we present an additional justification of the definition of community outliers we used and implemented in the \textbf{ABCD+o} algorithm, which assumes that outliers should have neighbors in various communities, while non-outliers should have neighbors concentrated in a single community. For this analysis, we selected a graph that has weak communities (as opposed to College Football Graph that has relatively strong communities).

For the test we consider the email-Eu graph built from email data from a European institution. In this dataset, taken from~\cite{snapnets}, an edge represents an email between two users. The data also has some ``ground-truth'' communities corresponding to 42 departments. Note, however, that most communities are ``very weak'' in the sense that there are more edges coming out of the community than within the community, as we discussed in Section~\ref{sec:participation}. Formally, we say that a set of nodes $C$ forms a community if 
$$
\sum_{v \in C} |N(v) \cap C| > \sum_{v \in C} |N(v) \setminus C|,
$$
where $N(v)$ denotes the set of neighbours of $v$. This property is \emph{not} satisfied for most communities in this dataset. Some communities are also very small: 16 communities have less than 10 nodes, and there are even some communities of size 1. We treat the graph as undirected and reduce it to its 2-core, which yields 934 nodes and 25,500 edges, with 580 nodes of degree 25 or more. 

Another way to measure the presence of community structure in a network is the modularity function, which is at the same time a quality function of many community detection algorithms. The definition of modularity for graphs was first introduced by Newman and Girvan in~\cite{newman2004finding}. It favors partitions of the set of nodes of a graph $G$ in which a large proportion of the edges falls entirely within the parts, but benchmarks it against the expected number of edges one would see in those parts in the corresponding \textbf{Chung-Lu} random graph model~\cite{chung2006complex} (the null model), which generates graphs with the expected degree sequence following exactly the degree sequence in $G$.

Formally, for a graph $G=(V,E)$ and a given partition $\A = \{A_1, A_2, \ldots, A_{\ell}\}$ of $V$, the \emph{modularity function} is defined as follows:
\begin{eqnarray}
q(\A) &=& \sum_{A_i \in \A} \frac{e(A_i)}{|E|}  - \sum_{A_i \in \A} \left( \frac{\vol(A_i)}{\vol(V)} \right)^2, \label{eq:q_G_A}
\end{eqnarray}
where for any $A \subseteq V$, $e(A)$ is the number of edges in the subgraph of $G$ \emph{induced by} set $A$, and $\vol(A) = \sum_{v \in A} \deg(v)$ is the \emph{volume} of set $A$. The first term in~(\ref{eq:q_G_A}), $\sum_{A_i \in \A} e(A_i)/|E|$, is called the \emph{edge contribution} and it computes the fraction of edges that fall within one of the parts. The second one, $\sum_{A_i \in \A} (\vol(A_i)/\vol(V))^2$, is called the \emph{degree tax} and it computes the expected fraction of edges that do the same in the corresponding random graph. The modularity measures the deviation between the two. Coming back to the dataset we use in our experiment, the modularity of the ground-truth communities (departments) is only $q_{gt} = 0.315$ even after reducing the network to its 2-core.

When applying clustering algorithms to such graphs with weak community structure, it is likely that the communities that are found are denser than the ``ground truth'' communities. This is due to the very nature of graph clustering algorithms which try to group nodes so that the corresponding communities are as dense as possible. Clustering this graph with \textbf{ECG} yields a smaller number of communities (33), many of which have a size less than 10 and the modularity is $q_{ECG} = 0.430$, larger than the one corresponding to the ground-truth communities. The communities found are also denser which can be seen by computing the edge contribution portion of the modularity function that measures the proportion of edges that fall within communities. This value is equal to 0.363 with the ground truth communities but climbs to 0.567 with \textbf{ECG} communities.

In practical applications ground-truth communities are often not known. This can especially be an issue if we work with graphs that have weak communities, such as the one we picked here. For this reason, we perform analysis of properties of nodes that are strongly identified as outliers against nodes that are strongly identified as non-outliers both against ground-truth communities and communities identified using community detection algorithm (which, as we discussed above for ``weak communities'' can significantly differ).

We consider two out of the four measures we introduced earlier to find outliers, namely, the \emph{ECG coefficient} (see Subsection~\ref{sec:ECG}), and the \emph{community association strength} (see Subsection~\ref{sec:edges}). We investigate the properties of the nodes with the highest (respectively lowest) scores, where low scores are indicative of outliers. In Table~\ref{tab:eu_low}, we plot some statistics for the nodes with a degree of 25 or more having small scores with respect to both measures, while in Table~\ref{tab:eu_hi}, we do the same for nodes with high scores. For each node, we look at (i) the proportion of edges in its own ECG community, (ii) the proportion of edges in its own ground-truth community, and (iii) the number of ground-truth communities in the neighborhood. For the first group of nodes, we clearly see that a minority of edges are internal to its ECG community, even more so if we look at the ground-truth communities. We also see that those nodes have neighbors is several different departments (ground-truth communities). The conclusions are exactly the opposite for the second group of nodes, as expected.

\begin{table}[]
\centering
\begin{tabular}{c | c | c}
~Prop. of edges in own~ &  ~Prop. of edges in own &  ~Number of ground-truth  \\
ECG community & ~ground-truth community~ & communities touched \\
\hline \hline
 0.223 &   0.052 &   34 \\
   0.145 &    0.048 &    32 \\
  0.275 &     0.076 &  31 \\
 0.143 &    0.037 &   36 \\
 0.185 &      0.222 &    31 \\
  0.079 &   0.059 &   29 \\
  0.217 &    0.137 &    26 \\
  0.159 &    0.038 &    29 \\
  0.167 &  0.056 &     27 \\
\end{tabular}
\vspace{.1cm}

\caption{Statistics for nodes with degree 25 or more and {\bf small} \emph{ECG coefficients} and \emph{community association strength} scores}
\label{tab:eu_low}
\end{table}

\begin{table}[]
\centering
\begin{tabular}{c | c | c}
~Prop. of edges in own~ &  ~Prop. of edges in own &  ~Number of ground-truth  \\
ECG community & ~ground-truth community~ & communities touched \\
\hline \hline
1.00 &     0.983 &      2 \\
1.00 &    1.000 &     1 \\
 0.97 &     0.970 &     2 \\
 1.00 &     0.981 &     2 \\
1.00 &    1.000 &      1 \\
 1.00 &     0.967 &  2 \\
   1.00 &   0.969 &   2 \\
  1.00 &    1.000 &    1 \\
\end{tabular}
\vspace{.1cm}

\caption{Statistics for nodes with degree 25 or more and {\bf large} \emph{ECG coefficients} and \emph{community association strength} scores}
\label{tab:eu_hi}
\end{table}

The email-Eu graph investigated in this subsection does not have outliers identified in its ground truth. Nevertheless, even with such a “noisy” graph, we find that nodes with the lowest scores are in contact with a very large number of distinct communities (Table~\ref{tab:eu_low}), while nodes with the highest scores are almost only in contact with other nodes in their own community (Table~\ref{tab:eu_hi}). This agrees with the concept of outlier which consists of nodes that appear to be making connections randomly to different communities, while non-outlier nodes make most of their connections within one or a small number of communities. These observations support the design assumptions behind the \textbf{ABCD+o} benchmark graph generator.


\section{Conclusions and Future Directions}\label{sec:future}

In this paper, we extended the \ABCD\ model to \textbf{ABCD+o} which incorporates the presence of outliers. We investigated selected properties that are able to distinguish outliers from regular nodes. We used two real-world graphs: College Football Graph and email-Eu graph, that are structurally significantly different, to both justify the design decisions behind \textbf{ABCD+o} generator and test the usefulness of this generator as a benchmark model. However, as a future direction, it would be valuable to perform more experiments with larger and topologically as well as structurally different networks to confirm our observations of which graph features are good predictors of outliers and which are not. After such verification, one may try to extend these ideas further and build an outlier detection algorithm and, in particular, use the \textbf{ABCD-o} benchmark we propose in this paper to validate it.


Another important extension of the original \ABCD\ model that we leave for the future is to design a variant of the model to include overlapping clusters. \textbf{ABCD+o} and the experience we gained by investigating properties of outliers are important stepping stones in that direction. Indeed, informally speaking, outliers are the nodes that do not strongly belong to any of the communities. But, clearly, one should distinguish a situation in which most of the neighbors of a given node belong to e.g.\ two communities from a situation in which neighbors are ``sprinkled'' across the entire graph. More refined properties may be able to extract information that is needed to distinguish the two scenarios and be used to build an unsupervised algorithm that is able to separate outliers from nodes that belong to multiple communities. With a better understanding of these properties, we should be able to adjust the \ABCD\ model one more time to incorporate both types of nodes.

An orthogonal future direction that we (and industry partners that we collaborate with) are interested in is to design a hypergraph model with known community structure and outliers. The first step is already made toward this goal~\cite{kaminski2022hypergraph}.

\section{Declarations}

\subsubsection*{Availability of data and materials}

The datasets generated and analyzed can be found in the associated Jupyter notebook and can be found on GitHub repository\footnote{\url{https://github.com/ftheberge/ABCDoExperiments}}.

\subsubsection*{Competing interests}

The authors declare that they have no competing interests.

\subsubsection*{Funding}

The research program of PP is partially supported by NSERC under Discovery Grant No.\ 2022-03804. Research of BK was supported by the Polish National Agency for Academic Exchange under the Strategic Partnerships programme, grant number BPI/PST/2021/1/00069/U/00001.

\subsubsection*{Authors' contributions}

All authors contributed equally. 

\subsubsection*{Acknowledgements}

Not applicable.

\bibliography{ref}

\begin{thebibliography}{10}

\bibitem{akoglu2015graph}
Leman Akoglu, Hanghang Tong, and Danai Koutra.
\newblock Graph based anomaly detection and description: a survey.
\newblock {\em Data mining and knowledge discovery}, 29(3):626--688, 2015.

\bibitem{bandyopadhyay2020integrating}
Sambaran Bandyopadhyay, Saley~Vishal Vivek, and M~Narasimha Murty.
\newblock Integrating network embedding and community outlier detection via
  multiclass graph description.
\newblock {\em arXiv preprint arXiv:2007.10231}, 2020.

\bibitem{bender1978asymptotic}
Edward~A Bender and E~Rodney Canfield.
\newblock The asymptotic number of labeled graphs with given degree sequences.
\newblock {\em Journal of Combinatorial Theory, Series A}, 24(3):296--307,
  1978.

\bibitem{bezanson2014}
Jeff Bezanson, Alan Edelman, Stefan Karpinski, and Viral~B. Shah.
\newblock Julia: A fresh approach to numerical computing, 2014.
\newblock URL: \url{https://arxiv.org/abs/1411.1607}, \href
  {https://doi.org/10.48550/ARXIV.1411.1607}
  {\path{doi:10.48550/ARXIV.1411.1607}}.

\bibitem{bollobas1980probabilistic}
B{\'e}la Bollob{\'a}s.
\newblock A probabilistic proof of an asymptotic formula for the number of
  labelled regular graphs.
\newblock {\em European Journal of Combinatorics}, 1(4):311--316, 1980.

\bibitem{Chakrabarti2004}
Deepayan Chakrabarti.
\newblock Autopart: Parameter-free graph partitioning and outlier detection.
\newblock In Jean-Fran{\c{c}}ois Boulicaut, Floriana Esposito, Fosca Giannotti,
  and Dino Pedreschi, editors, {\em Knowledge Discovery in Databases: PKDD
  2004}, pages 112--124, Berlin, Heidelberg, 2004. Springer Berlin Heidelberg.

\bibitem{chung2006complex}
Fan Chung~Graham and Linyuan Lu.
\newblock {\em Complex graphs and networks}, volume 107 of {\em CBMS Regional
  Conference Series in Mathematics}.
\newblock American Mathematical Soc., 2006.

\bibitem{curado2023novel}
Manuel Curado, Leandro Tortosa, and Jose~F Vicent.
\newblock A novel measure to identify influential nodes: Return random walk
  gravity centrality.
\newblock {\em Information Sciences}, 2023.

\bibitem{flake2000efficient}
Gary~William Flake, Steve Lawrence, and C~Lee Giles.
\newblock Efficient identification of web communities.
\newblock In {\em Proceedings of the sixth ACM SIGKDD international conference
  on Knowledge discovery and data mining}, pages 150--160, 2000.

\bibitem{fortunato2010community}
Santo Fortunato.
\newblock Community detection in graphs.
\newblock {\em Physics reports}, 486(3-5):75--174, 2010.

\bibitem{gaucher2021outlier}
Solenne Gaucher, Olga Klopp, and Genevi{\`e}ve Robin.
\newblock Outlier detection in networks with missing links.
\newblock {\em Computational Statistics \& Data Analysis}, 164:107308, 2021.

\bibitem{ghalmane2019immunization}
Zakariya Ghalmane, Mohammed~El Hassouni, and Hocine Cherifi.
\newblock Immunization of networks with non-overlapping community structure.
\newblock {\em Social Network Analysis and Mining}, 9:1--22, 2019.

\bibitem{girvan2002community}
Michelle Girvan and Mark~EJ Newman.
\newblock Community structure in social and biological networks.
\newblock {\em Proceedings of the national academy of sciences},
  99(12):7821--7826, 2002.

\bibitem{gregory2010finding}
Steve Gregory.
\newblock Finding overlapping communities in networks by label propagation.
\newblock {\em New journal of Physics}, 12(10):103018, 2010.

\bibitem{grover2016node2vec}
Aditya Grover and Jure Leskovec.
\newblock node2vec: Scalable feature learning for networks.
\newblock In {\em Proceedings of the 22nd ACM SIGKDD international conference
  on Knowledge discovery and data mining}, pages 855--864, 2016.

\bibitem{javed2018community}
Muhammad~Aqib Javed, Muhammad~Shahzad Younis, Siddique Latif, Junaid Qadir, and
  Adeel Baig.
\newblock Community detection in networks: A multidisciplinary review.
\newblock {\em Journal of Network and Computer Applications}, 108:87--111,
  2018.

\bibitem{kaminski2021multi}
Bogumi{\l} Kami{\'n}ski, {\L}ukasz Krai{\'n}ski, Pawe{\l} Pra{\l}at, and
  Fran{\c{c}}ois Th{\'e}berge.
\newblock A multi-purposed unsupervised framework for comparing embeddings of
  undirected and directed graphs.
\newblock {\em Network Science}, 10(4):323--346, 2022.

\bibitem{kaminski2022fast}
Bogumi{\l} Kami{\'n}ski, Tomasz Olczak, Bartosz Pankratz, Pawe{\l} Pra{\l}at,
  and Fran{\c{c}}ois Th{\'e}berge.
\newblock Properties and performance of the abcde random graph model with
  community structure.
\newblock {\em Big Data Research}, 30:100348, 2022.

\bibitem{kaminski2022modularity}
Bogumi{\l} Kami{\'n}ski, Bartosz Pankratz, Pawe{\l} Pra{\l}at, and
  Fran{\c{c}}ois Th{\'e}berge.
\newblock Modularity of the abcd random graph model with community structure.
\newblock {\em Journal of Complex Networks}, 10(6):cnac050, 2022.

\bibitem{kaminski2022outliers}
Bogumi{\l} Kami{\'n}ski, Pawe{\l} Pra{\l}at, and Th{\'e}berge.
\newblock Outliers in the abcd random graph model with community structure
  (abcd+o).
\newblock In {\em 11th International Conference on Complex Networks and Their
  Applications}. Springer Studies in Computational Intelligence (SCI, volume
  1078), 2022.

\bibitem{kaminski2020unsupervised}
Bogumi{\l} Kami{\'n}ski, Pawe{\l} Pra{\l}at, and Fran{\c{c}}ois Th{\'e}berge.
\newblock An unsupervised framework for comparing graph embeddings.
\newblock {\em Journal of Complex Networks}, 8(5):cnz043, 2020.

\bibitem{kaminski2021artificial}
Bogumi{\l} Kami{\'n}ski, Pawe{\l} Pra{\l}at, and Fran{\c{c}}ois Th{\'e}berge.
\newblock Artificial benchmark for community detection (abcd)—fast random
  graph model with community structure.
\newblock {\em Network Science}, pages 1--26, 2021.

\bibitem{kaminski2021mining}
Bogumi{\l} Kami{\'n}ski, Pawe{\l} Pra{\l}at, and Fran{\c{c}}ois Th{\'e}berge.
\newblock {\em Mining Complex Networks}.
\newblock CRC Press, 2021.

\bibitem{kaminski2022hypergraph}
Bogumi{\l} Kami{\'n}ski, Pawe{\l} Pra{\l}at, and Fran{\c{c}}ois Th{\'e}berge.
\newblock Hypergraph artificial benchmark for community detection (h-abcd).
\newblock {\em arXiv preprint arXiv:2210.15009}, 2022.

\bibitem{lancichinetti2009benchmarks}
Andrea Lancichinetti and Santo Fortunato.
\newblock Benchmarks for testing community detection algorithms on directed and
  weighted graphs with overlapping communities.
\newblock {\em Physical Review E}, 80(1):016118, 2009.

\bibitem{lancichinetti2008benchmark}
Andrea Lancichinetti, Santo Fortunato, and Filippo Radicchi.
\newblock Benchmark graphs for testing community detection algorithms.
\newblock {\em Physical review E}, 78(4):046110, 2008.

\bibitem{snapnets}
Jure Leskovec and Andrej Krevl.
\newblock {SNAP Datasets}: {Stanford} large network dataset collection.
\newblock \url{http://snap.stanford.edu/data}, June 2014.

\bibitem{Liu2015}
Dehai Liu, Benjin Mei, Jinchuan Chen, Zhiwu Lu, and Xiaoyong Du.
\newblock Community based spammer detection in social networks.
\newblock In Xin~Luna Dong, Xiaohui Yu, Jian Li, and Yizhou Sun, editors, {\em
  Web-Age Information Management}, pages 554--558, Cham, 2015. Springer
  International Publishing.

\bibitem{liu2020gmm}
Fan Liu, Zhen Wang, and Yong Deng.
\newblock Gmm: A generalized mechanics model for identifying the importance of
  nodes in complex networks.
\newblock {\em Knowledge-Based Systems}, 193:105464, 2020.

\bibitem{lu2018community}
Zhenqi Lu, Johan Wahlstr{\"o}m, and Arye Nehorai.
\newblock Community detection in complex networks via clique conductance.
\newblock {\em Scientific reports}, 8(1):1--16, 2018.

\bibitem{Newman2018}
M.~E.~J. Newman.
\newblock {\em Networks (2nd edition)}.
\newblock Oxford University Press, Oxford; New York, 2018.

\bibitem{newman2004finding}
Mark~EJ Newman and Michelle Girvan.
\newblock Finding and evaluating community structure in networks.
\newblock {\em Physical review E}, 69(2):026113, 2004.

\bibitem{poulin2018ensemble}
Val{\'e}rie Poulin and Fran{\c{c}}ois Th{\'e}berge.
\newblock Ensemble clustering for graphs.
\newblock In {\em International Conference on Complex Networks and their
  Applications}, pages 231--243. Springer, 2018.

\bibitem{radicchi2004defining}
Filippo Radicchi, Claudio Castellano, Federico Cecconi, Vittorio Loreto, and
  Domenico Parisi.
\newblock Defining and identifying communities in networks.
\newblock {\em Proceedings of the national academy of sciences},
  101(9):2658--2663, 2004.

\bibitem{rajeh2023comparative}
Stephany Rajeh, Marinette Savonnet, Eric Leclercq, and Hocine Cherifi.
\newblock Comparative evaluation of community-aware centrality measures.
\newblock {\em Quality \& Quantity}, 57(2):1273--1302, 2023.

\bibitem{singh2021ni}
Dipika Singh and Rakhi Garg.
\newblock Ni-louvain: A novel algorithm to detect overlapping communities with
  influence analysis.
\newblock {\em Journal of King Saud University-Computer and Information
  Sciences}, 2021.

\bibitem{sun2005}
Jimeng Sun, Huiming Qu, D.~Chakrabarti, and C.~Faloutsos.
\newblock Neighborhood formation and anomaly detection in bipartite graphs.
\newblock In {\em Fifth IEEE International Conference on Data Mining
  (ICDM'05)}, pages 8 pp.--, 2005.
\newblock \href {https://doi.org/10.1109/ICDM.2005.103}
  {\path{doi:10.1109/ICDM.2005.103}}.

\bibitem{Viswanath2010}
Bimal Viswanath, Ansley Post, Krishna~P. Gummadi, and Alan Mislove.
\newblock An analysis of social network-based sybil defenses.
\newblock {\em SIGCOMM Comput. Commun. Rev.}, 40(4):363–374, aug 2010.
\newblock \href {https://doi.org/10.1145/1851275.1851226}
  {\path{doi:10.1145/1851275.1851226}}.

\bibitem{wormald1984generating}
Nicholas~C Wormald.
\newblock Generating random regular graphs.
\newblock {\em Journal of algorithms}, 5(2):247--280, 1984.

\bibitem{wormald1999models}
Nicholas~C Wormald et~al.
\newblock Models of random regular graphs.
\newblock {\em London Mathematical Society Lecture Note Series}, pages
  239--298, 1999.

\bibitem{yang2013overlapping}
Jaewon Yang and Jure Leskovec.
\newblock Overlapping community detection at scale: a nonnegative matrix
  factorization approach.
\newblock In {\em Proceedings of the sixth ACM international conference on Web
  search and data mining}, pages 587--596, 2013.

\end{thebibliography}

\appendix

\section{Node Properties---Plots Associated with Subsection~\ref{sec:properties}}

\begin{figure}[h]
\centering
\includegraphics[scale=0.53]{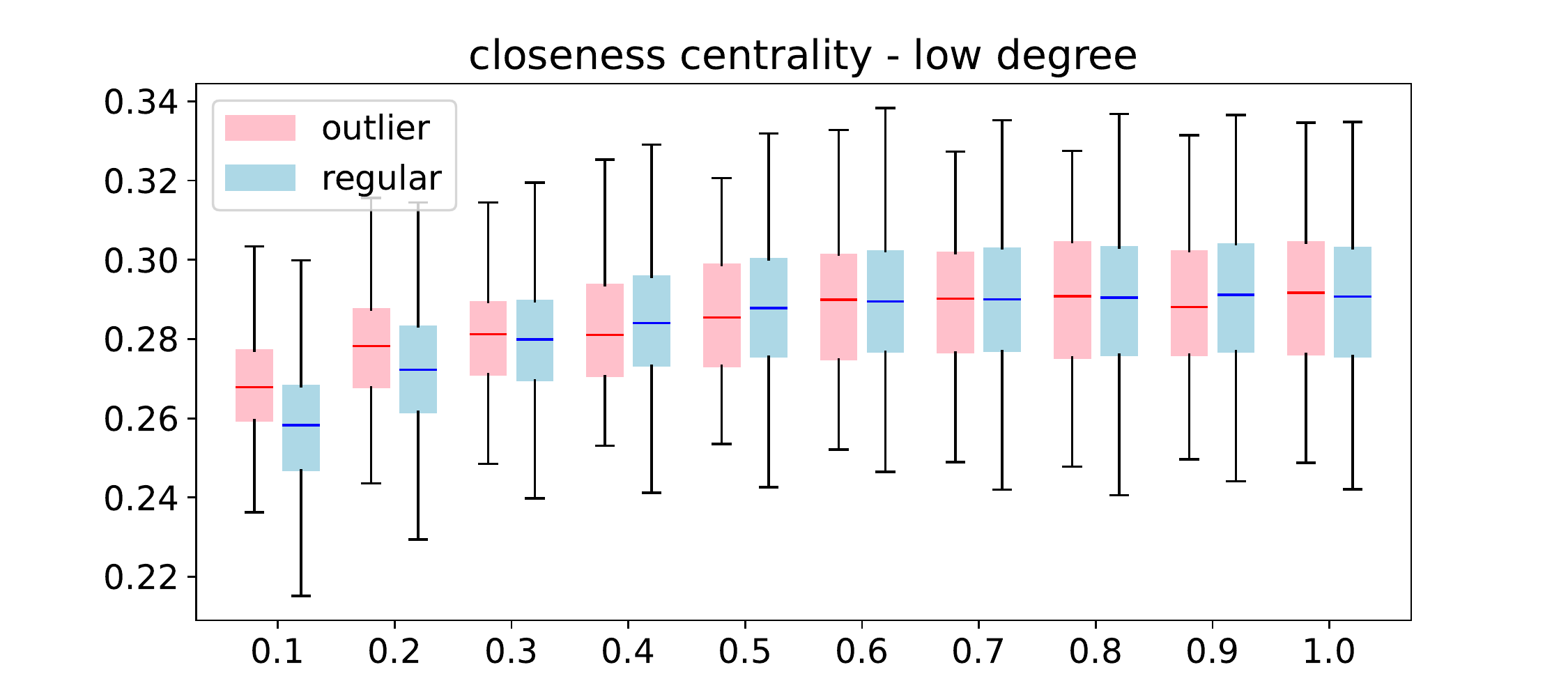}
\includegraphics[scale=0.53]{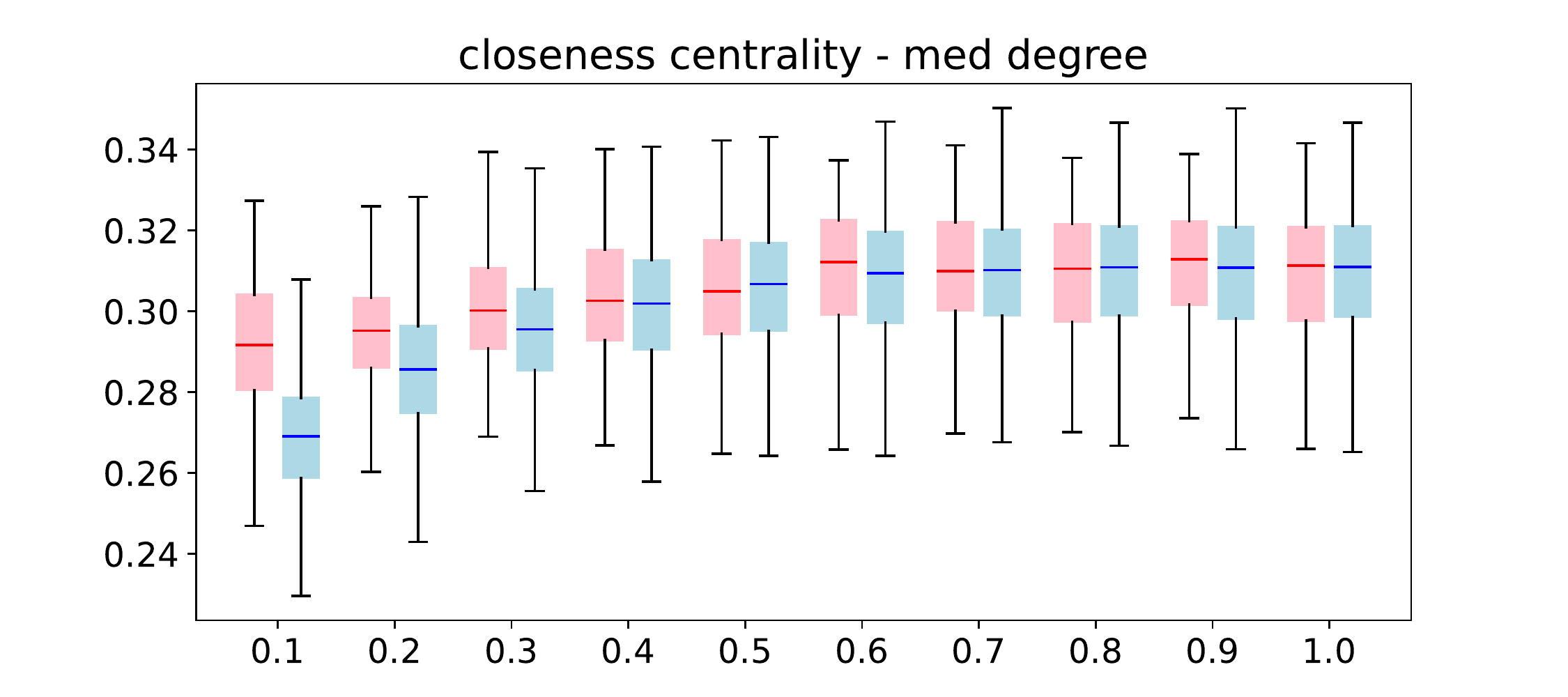}
\includegraphics[scale=0.53]{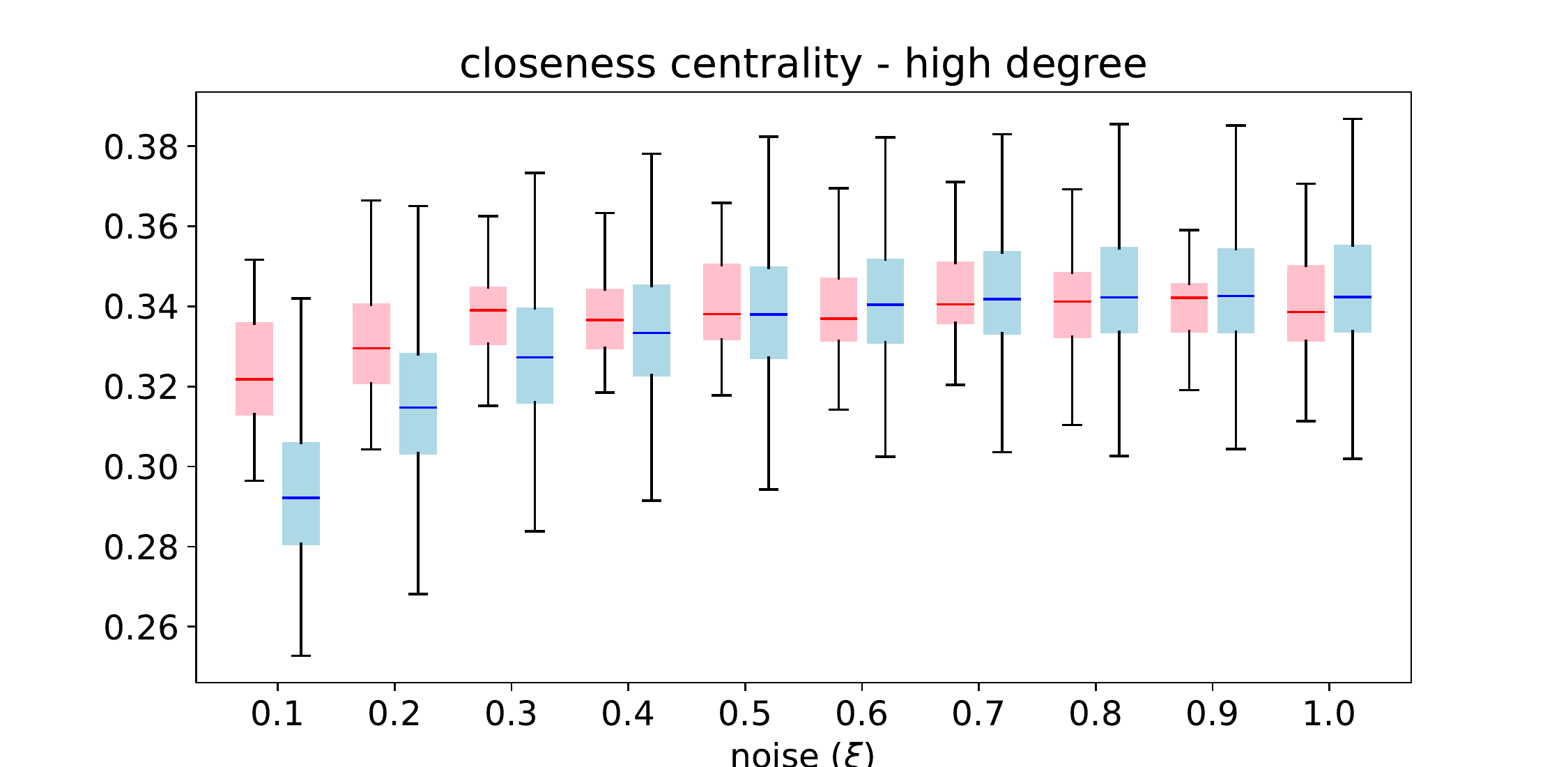}
\caption{Comparing closeness centrality of outlier and regular nodes for the \textbf{ABCD+o} graphs, respectively, for low, medium, and high degree nodes.}
\label{fig:close}
\end{figure}

\begin{figure}
\centering
\includegraphics[scale=0.53]{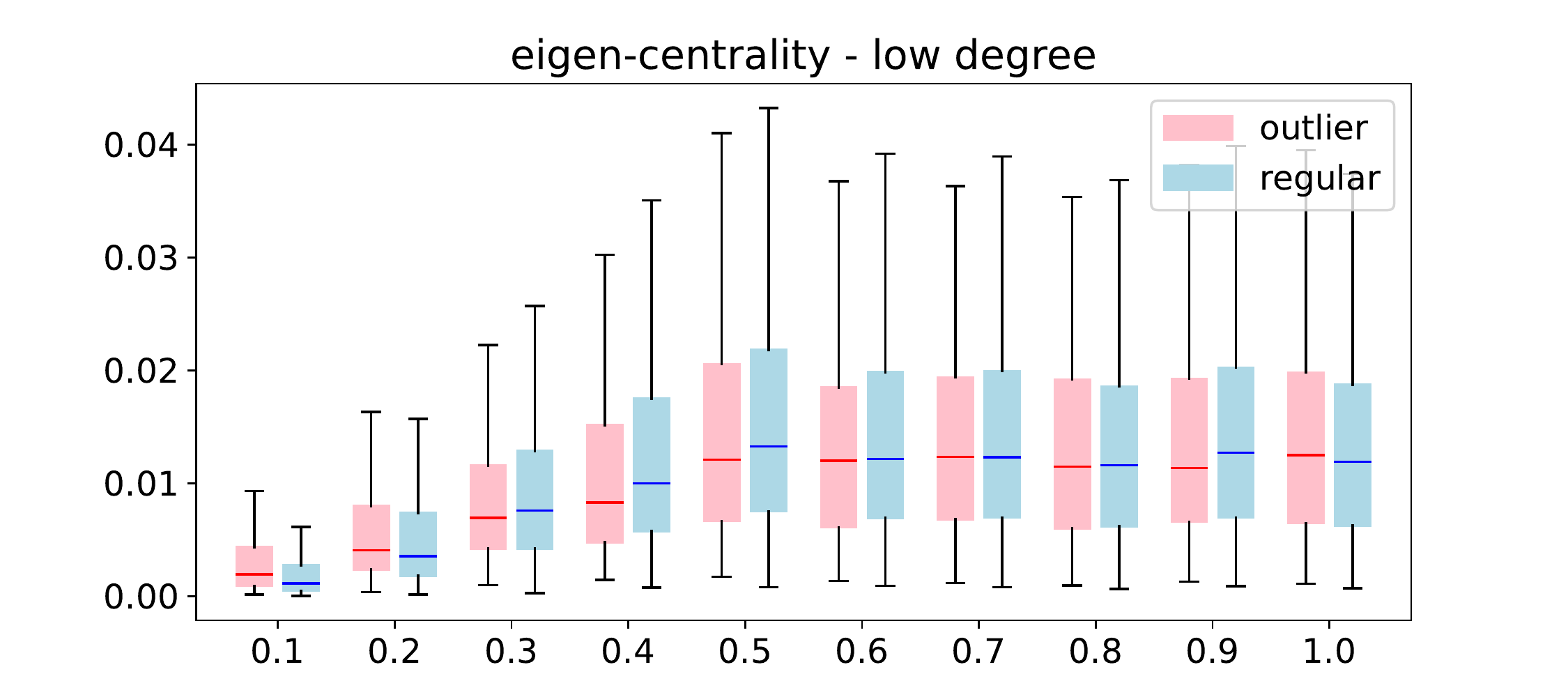}
\includegraphics[scale=0.53]{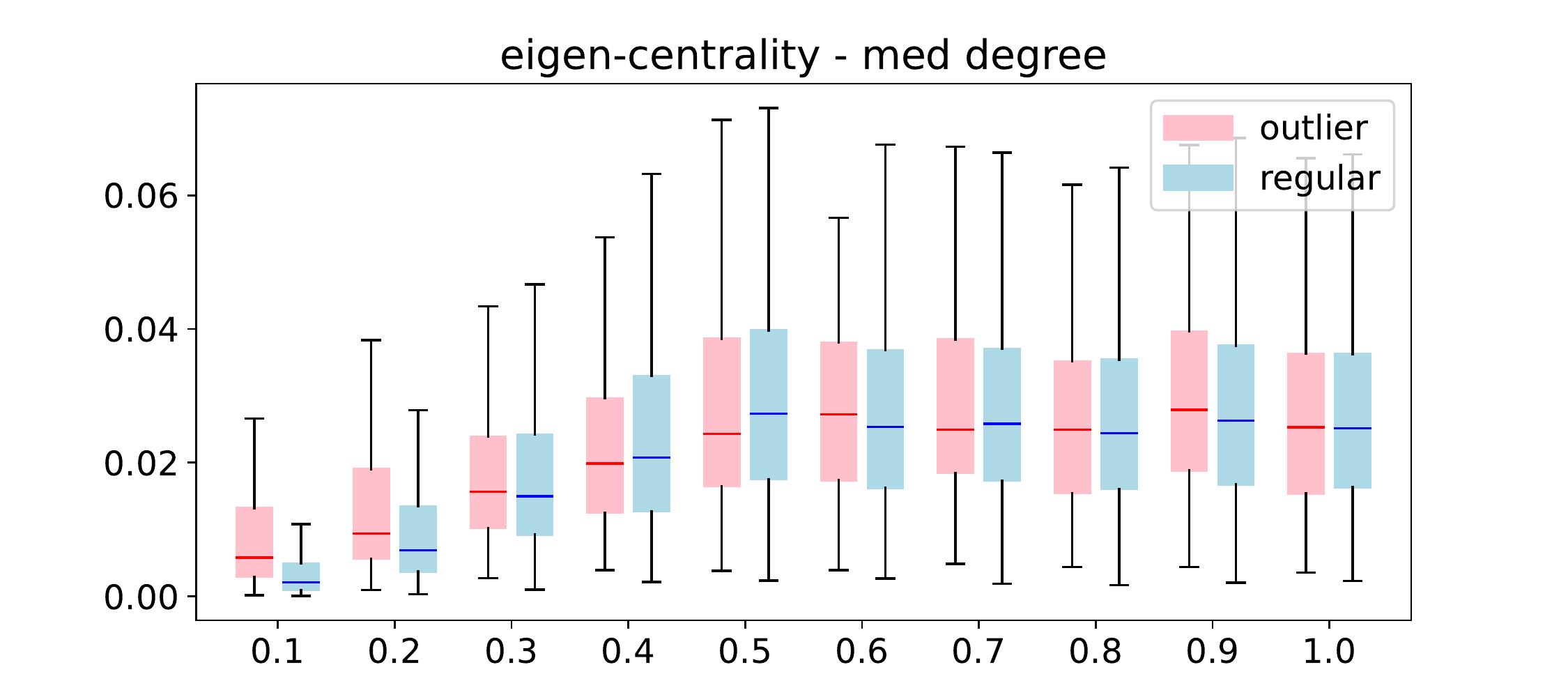}
\includegraphics[scale=0.53]{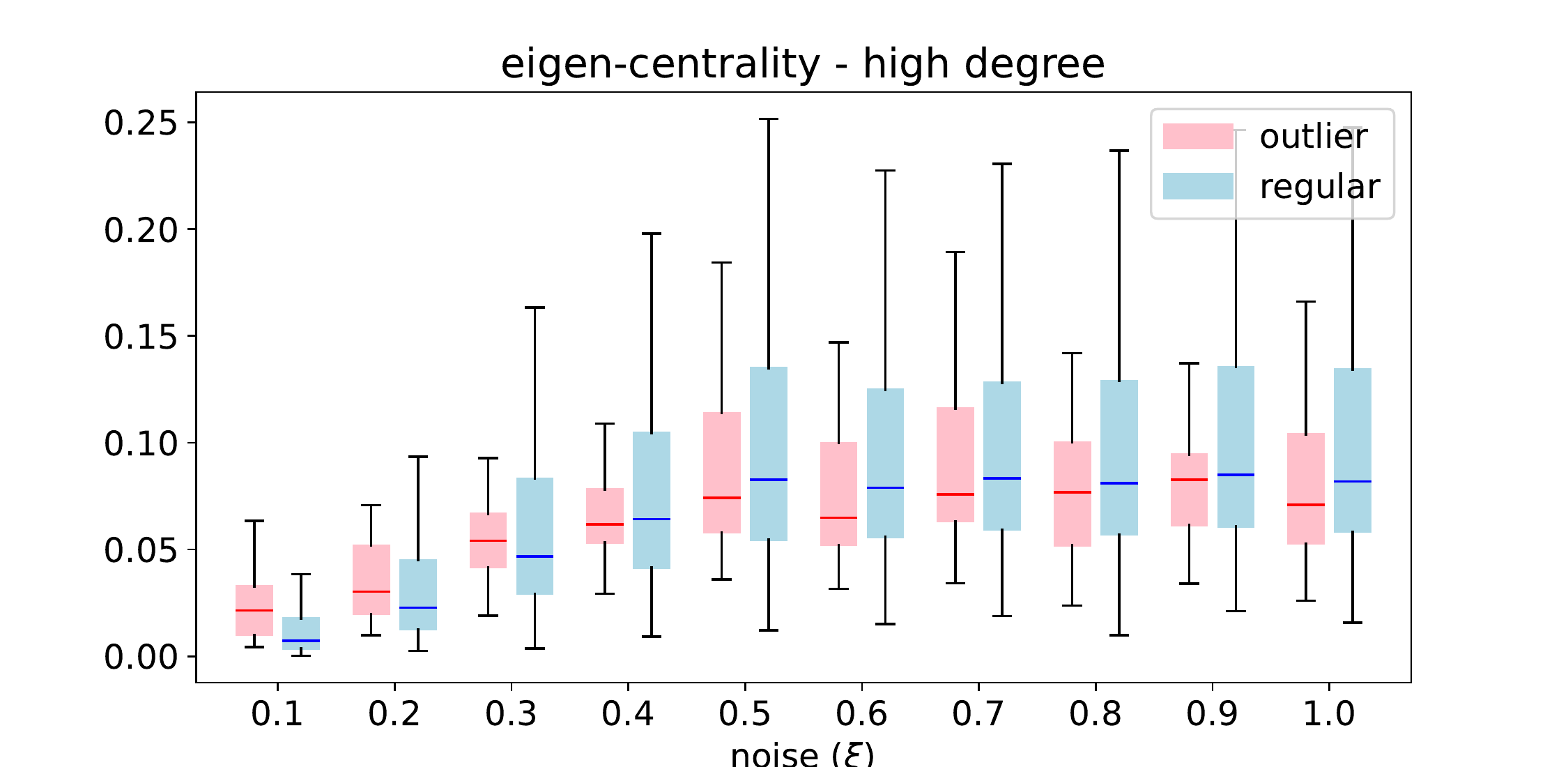}
\caption{Comparing eigen-centrality of outlier and regular nodes for the \textbf{ABCD+o} graphs, respectively, for low, medium, and high degree nodes.}
\label{fig:eigen}
\end{figure}

\begin{figure}
\centering
\includegraphics[scale=0.53]{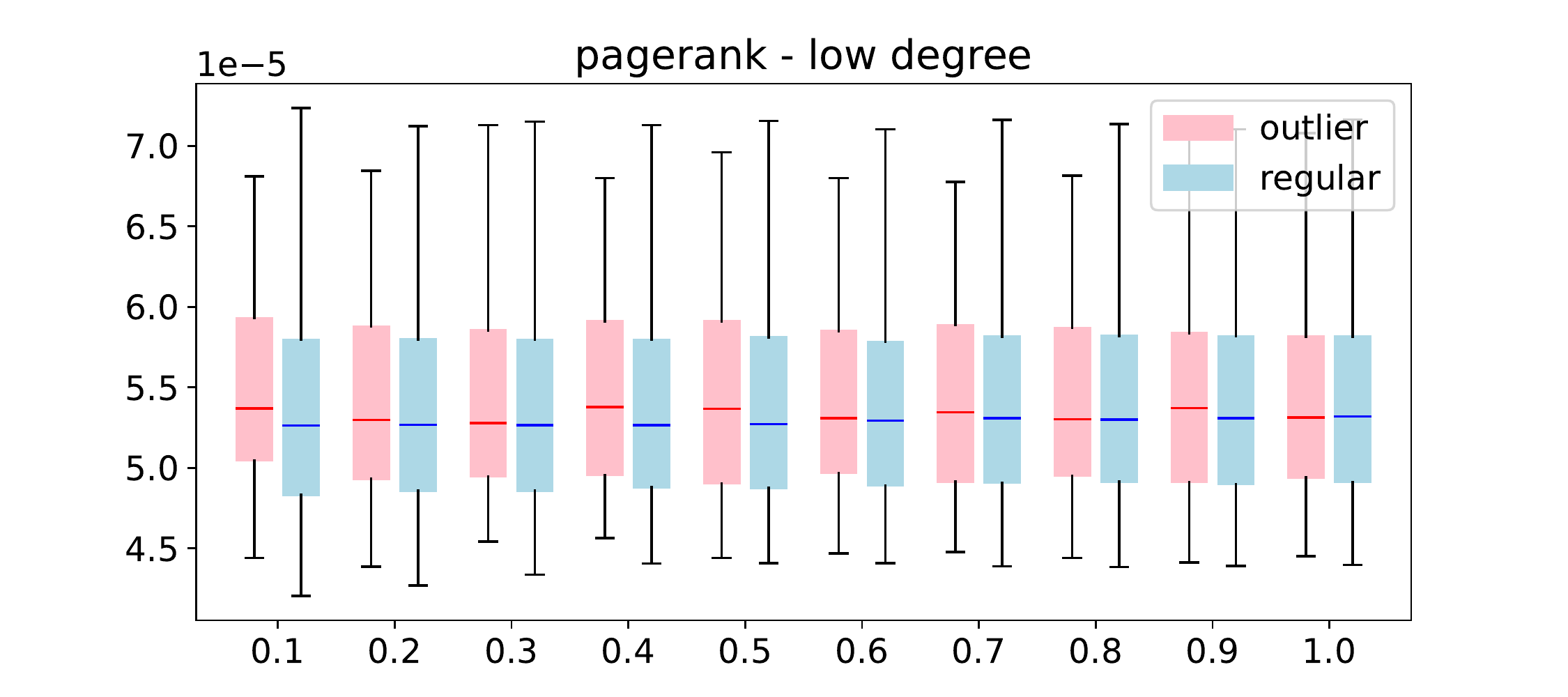}
\includegraphics[scale=0.53]{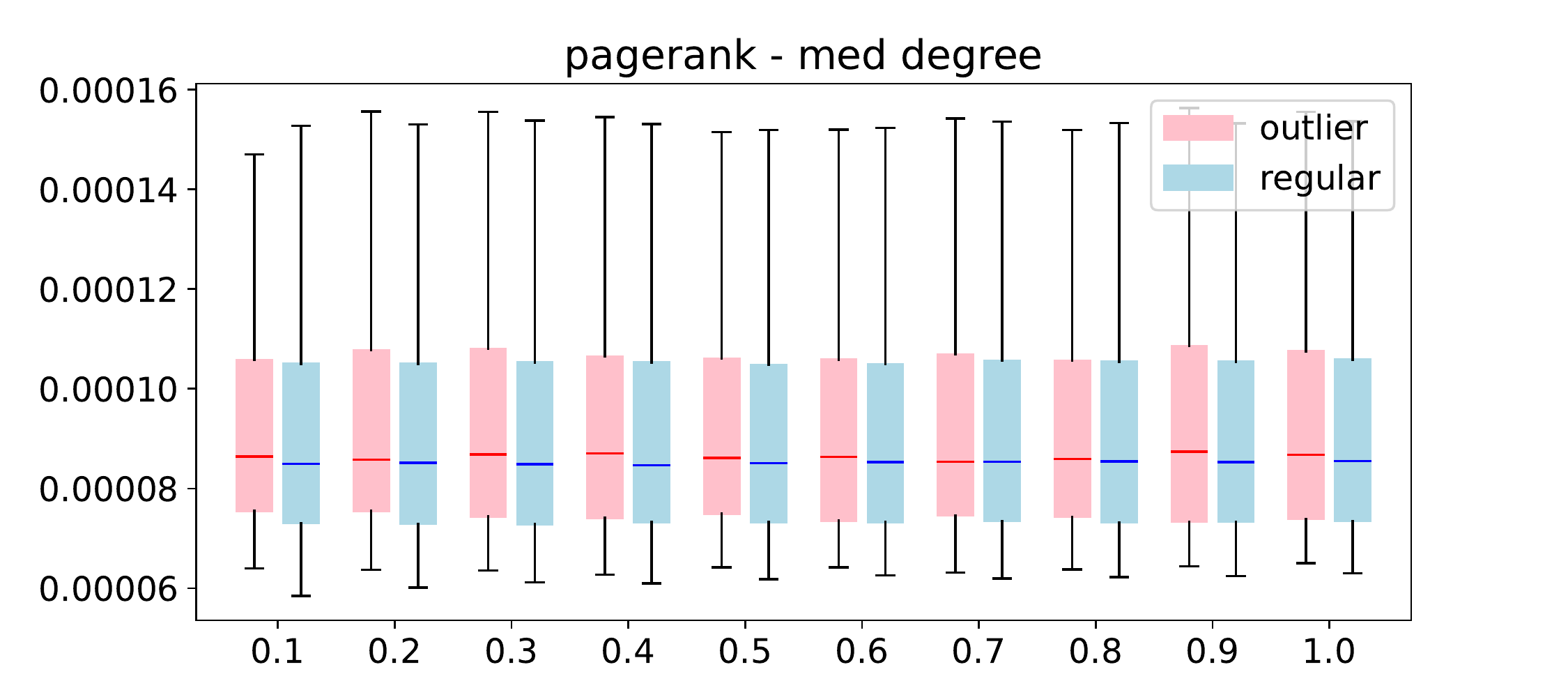}
\includegraphics[scale=0.53]{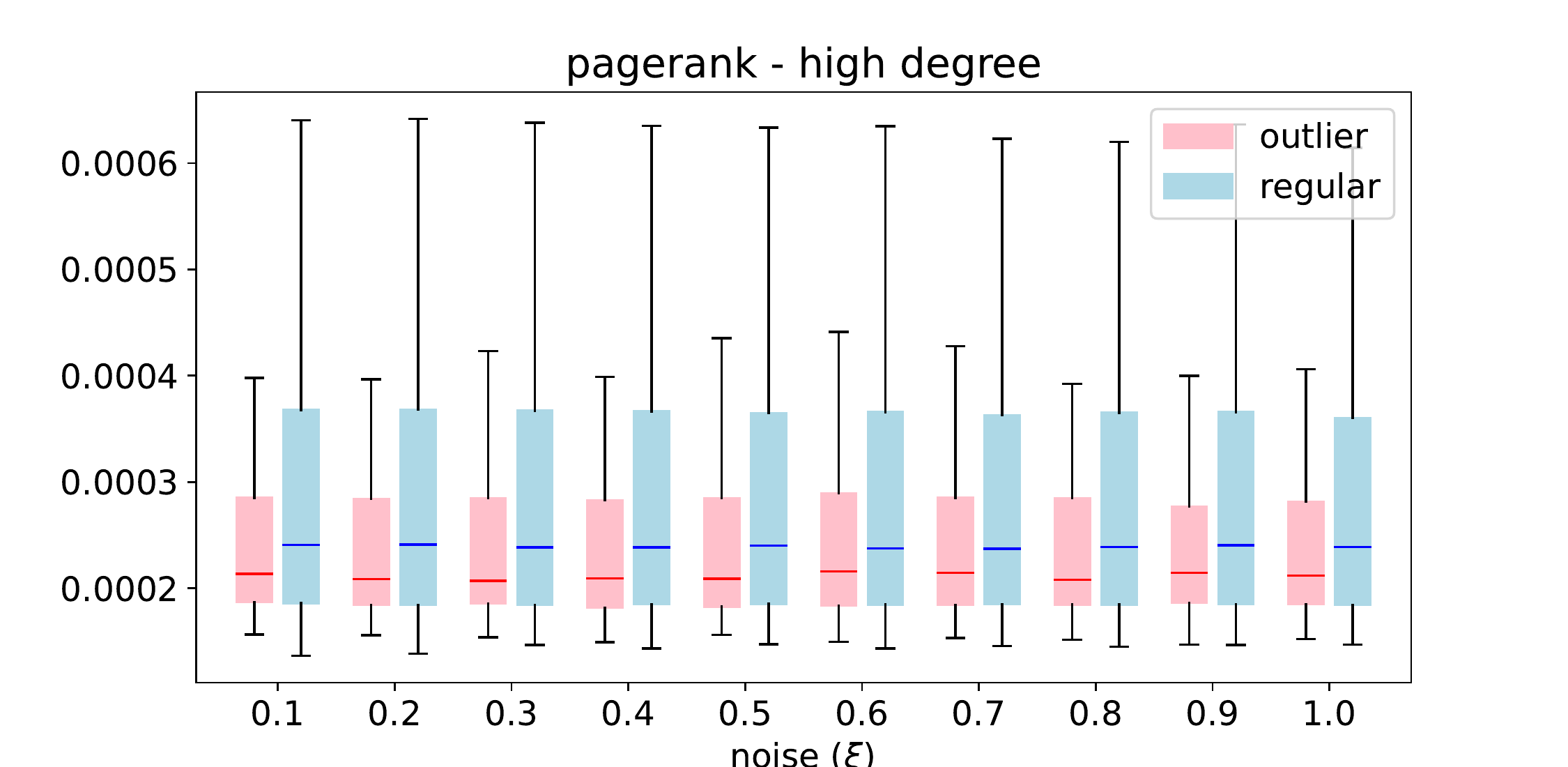}
\caption{Comparing pagerank scores of outlier and regular nodes for the \textbf{ABCD+o} graphs, respectively, for low, medium, and high degree nodes.}
\label{fig:page}
\end{figure}

\begin{figure}
\centering
\includegraphics[scale=0.53]{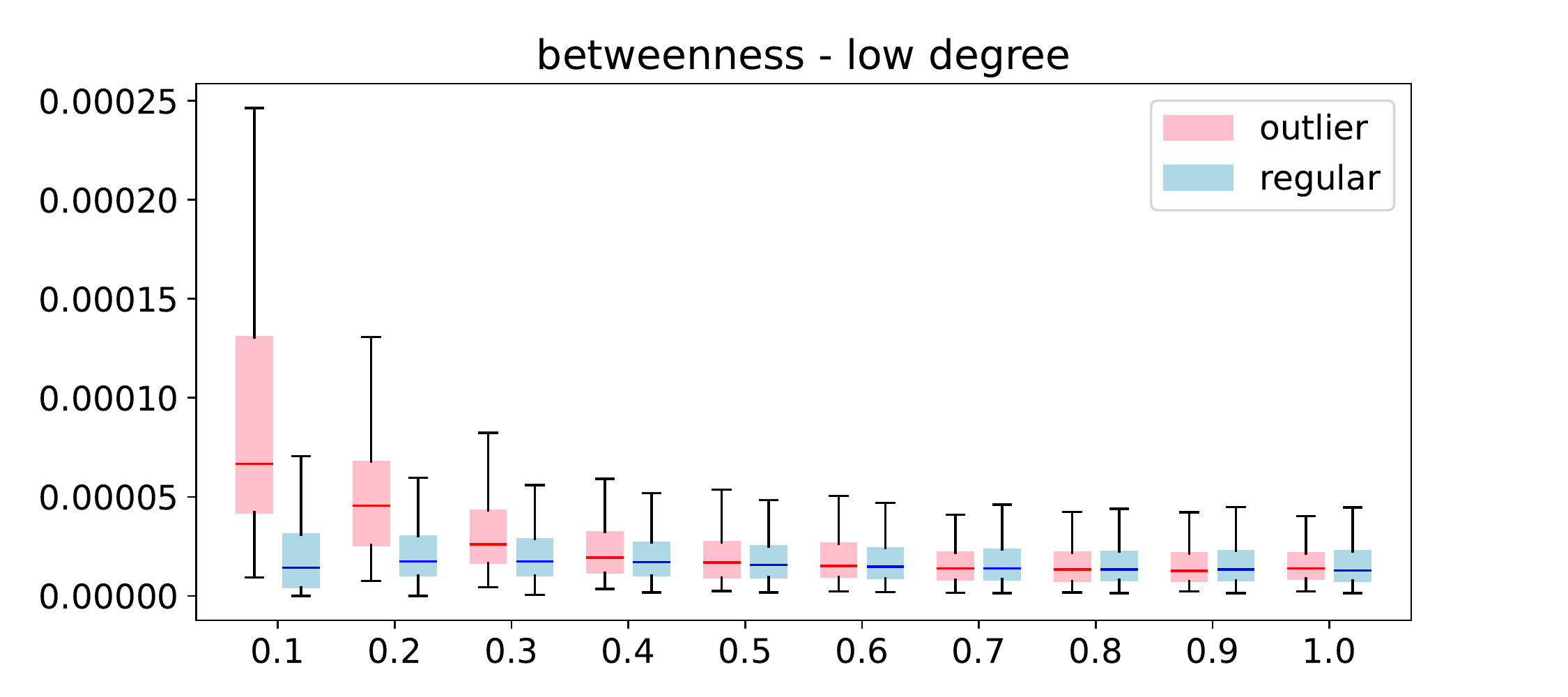}
\includegraphics[scale=0.53]{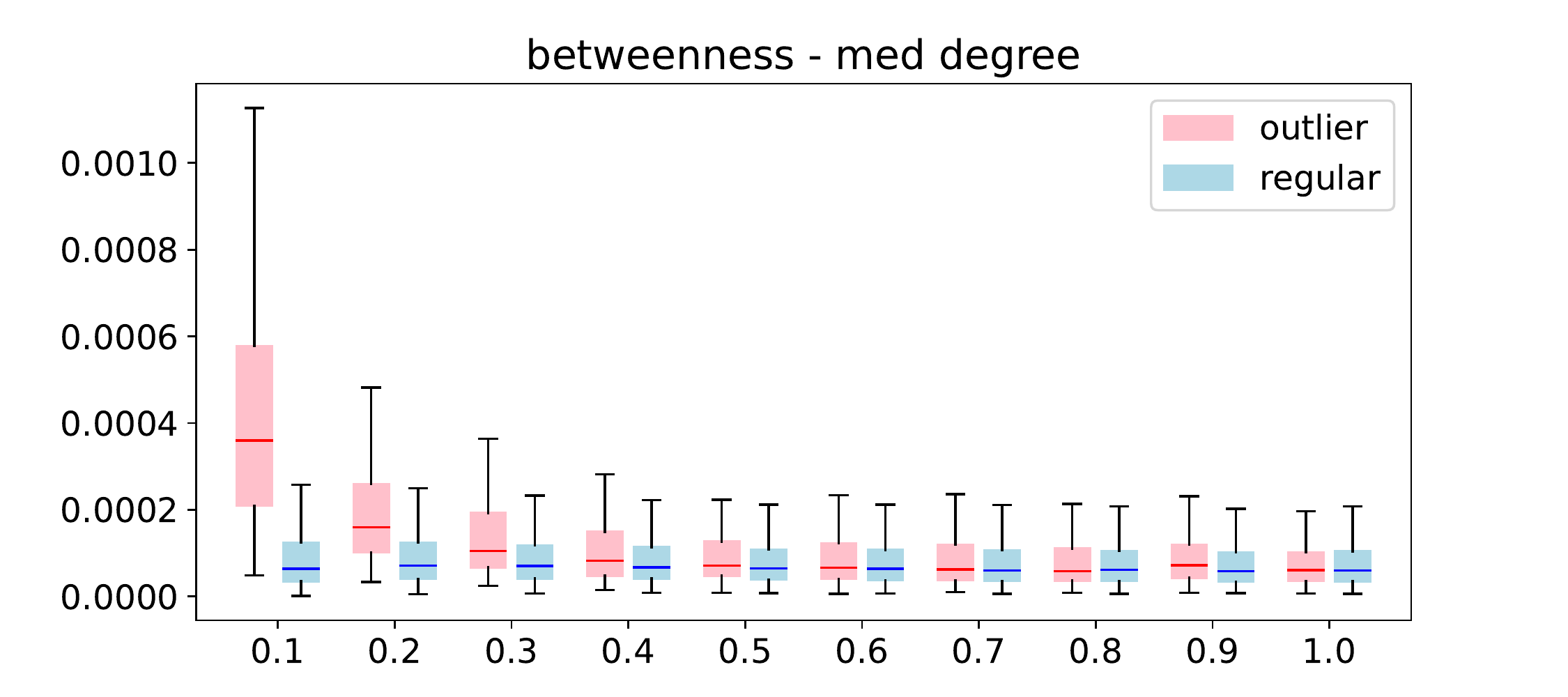}
\includegraphics[scale=0.53]{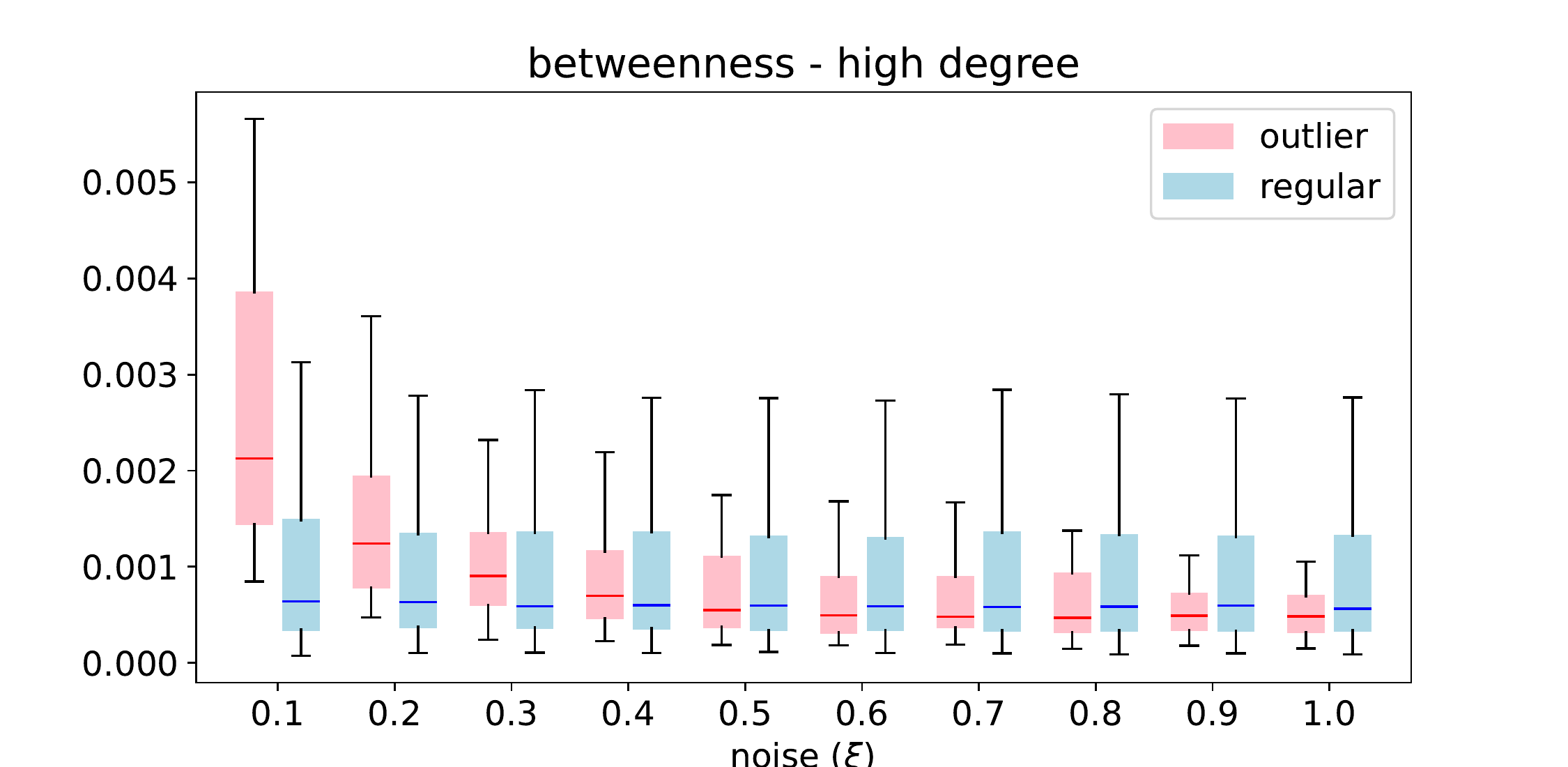}
\caption{Comparing betweenness scores of outlier and regular nodes for the \textbf{ABCD+o} graphs, respectively, for low, medium, and high degree nodes.}
\label{fig:between}
\end{figure}

\begin{figure}
\centering
\includegraphics[scale=0.48]{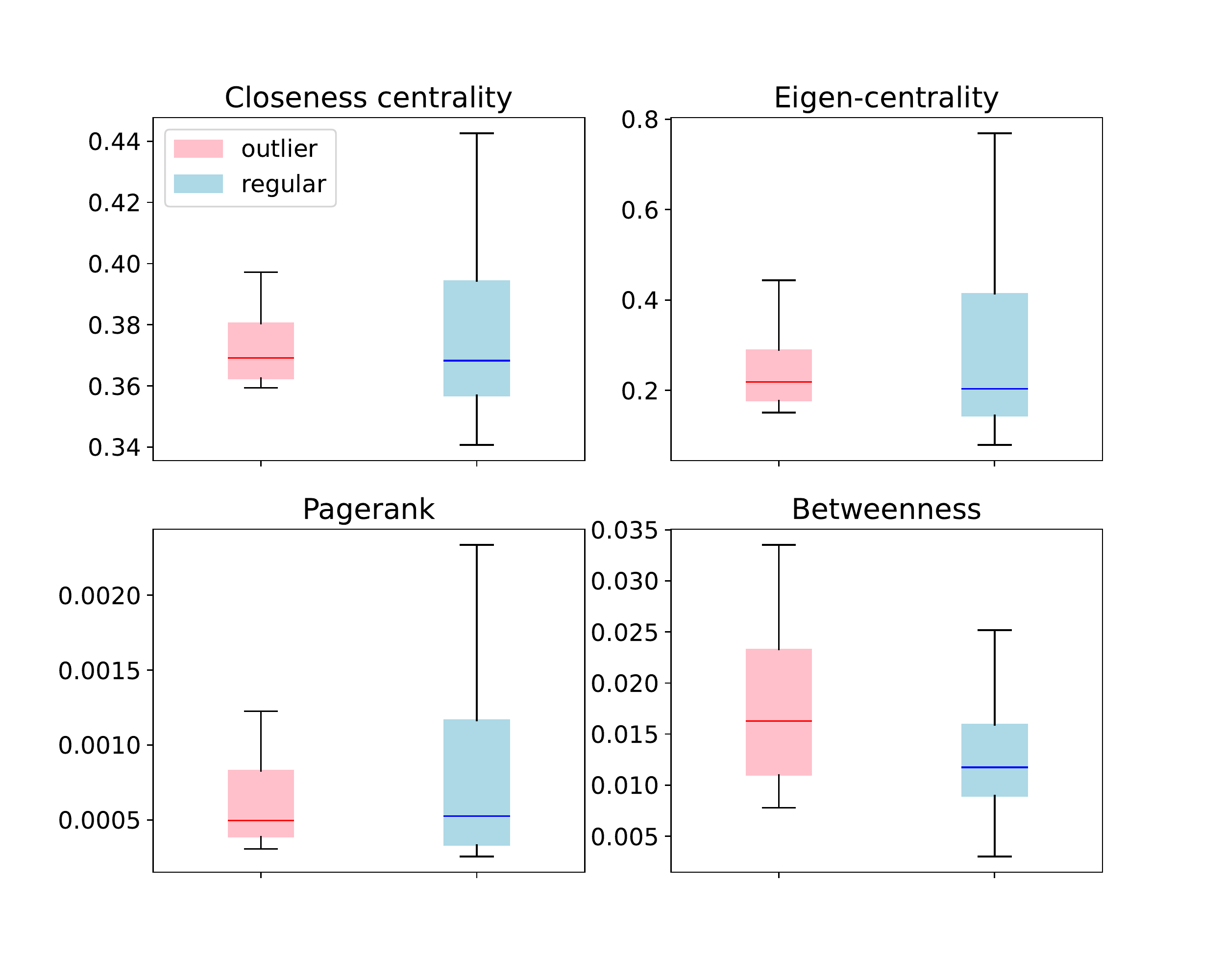}
\caption{Comparing centrality measures of outlier and regular nodes for the College Football Graph.}
\label{fig:foot_features}
\end{figure}

In Figures~\ref{fig:close} to~\ref{fig:between}, we compare the distribution of four centrality measures between outlier and non-outlier nodes: closeness centrality, eigen-centrality, PageRank, and betweenness. In each case, we show three plots looking at nodes with a low, a medium, and a high degree, respectively. We show results for the same measures for the College Football Graph in Figure~\ref{fig:foot_features}.

\end{document}